\documentclass[aps,prl,reprint,superscriptaddress,nobibnotes]{revtex4-2}

\usepackage{physics}
\usepackage{amssymb}
\usepackage{amsfonts}
\usepackage{dsfont}
\usepackage{hyperref}
\usepackage{soul}
\usepackage{tabularx}
\usepackage{graphicx}
\usepackage{xcolor}
\usepackage{makecell}
\usepackage{mathtools}

\newcommand{\normord}[1]{:\mathrel{#1}:}

\begin{document}

\title{Wavefunctions for Anyon Superconductors}

\author{Donghae Seo}
\thanks{These authors contributed equally.}
\affiliation{Natural Science Research Institute, Korea Advanced Institute of Science and Technology, Daejeon 34141, Republic of Korea}

\author{Taegon Lee}
\thanks{These authors contributed equally.}
\affiliation{Department of Physics, Korea Advanced Institute of Science and Technology, Daejeon 34141, Republic of Korea}

\author{Gil Young Cho}
\email{gilyoungcho@kaist.ac.kr}
\affiliation{Department of Physics, Korea Advanced Institute of Science and Technology, Daejeon 34141, Republic of Korea}
\affiliation{Center for Artificial Low Dimensional Electronic Systems, Institute for Basic Science, Pohang 37673, Republic of Korea}

\begin{abstract}
    Anyon superconductivity arises from the condensation of mobile anyons rather than from a conventional Cooper instability, yet a systematic wavefunction description remains lacking. We develop a hierarchy-wavefunction construction for superconducting states derived from parent topological orders and identify their off-diagonal long-range order, condensate charge, chiral central charge, and residual topological order through the plasma analogy and topological field theories. We construct Abelian and non-Abelian examples descending from the semion {state}, a $\nu=2/3$ hierarchy state, the $\nu=1/3$ Laughlin state, $\nu=1$ integer quantum Hall state, and Pfaffian state. Remarkably, for the semion case, the superconducting many-semion wavefunction is equivalent to a state of fermionized anyons filling two effective Landau levels, recovering Laughlin's {original} construction {of semion superconductor}. Finally, we show that hierarchy wavefunctions emerge naturally in the dilute, long-distance limit of the anyon-Hilbert-space formulation, which applies to ideal Chern bands and moir\'e bands of twisted bilayer MoTe\textsubscript{2}. Our results establish a unified wavefunction-level framework linking anyon condensation, superconducting order, and topological field theory.
\end{abstract}

\maketitle

Anyons are emergent quasiparticles that constitute a defining hallmark of long-range entanglement in two-dimensional topological orders~\cite{wen2017colloquium}. Their fractional statistics distinguish them from bosons and fermions, offering a route toward topologically protected quantum computation \cite{nayak2008non-abelian}. The recent realization of fractional Chern insulators in moir\'e materials has brought anyons into a highly tunable solid-state setting and stimulated interest in the proximate phases that can emerge from fractional Chern insulators \cite{cai2023signatures,zeng2023thermodynamic,park2023observation,xu2023observation,lu2024fractional}. Of particular interest is anyon superconductivity, in which superconducting coherence originates from the condensation of doped anyons, rather than from a conventional Cooper instability of electrons \cite{laughlin1988superconducting,laughlin1988relationship,fetter1989random-phase,chen1989anyon,lee1989anyon,wen1989chiral,hosotani1990superconductivity,wen1990compressibility,kitazawa1990exactness,fradkin1990superfluidity,lee1991anyon,shi2025doping,shi2025dopinglattice,divic2025anyon,pichler2026microscopic,nosov2026anyon,nakajima2026thermodynamics,guerci2025from,shi2025nonabelian,schleith2025anyon,zhang2025holon,seo2026unified,lotric2026phases,fan2026hidden,wang2026topological,shi2026superconductivity,zhang2026color,mehta2026coloring}. First proposed by Laughlin \cite{laughlin1988superconducting,laughlin1988relationship}, this mechanism has attracted renewed attention following the observation of superconductivity in twisted bilayer MoTe\textsubscript{2} \cite{xu2025signatures}. Although recent studies have developed field-theoretic and categorical descriptions of anyon superconductors \cite{shi2025doping,shi2025dopinglattice,divic2025anyon,pichler2026microscopic,nosov2026anyon,nakajima2026thermodynamics,guerci2025from,shi2025nonabelian,schleith2025anyon,zhang2025holon,seo2026unified,lotric2026phases,fan2026hidden,wang2026topological,shi2026superconductivity,zhang2026color,mehta2026coloring}, a systematic connection between these approaches and explicit many-body wavefunctions remains lacking, as does a direct wavefunction-level characterization of superconductivity. 

In this Letter, we present a systematic construction of wavefunctions for anyon superconductors derived from a parent topological order, motivated by the correspondence between quantum Hall hierarchy states and anyon superconductivity \cite{seo2026unified}. At a heuristic level, this connection is already suggestive: in both settings, doping a parent fractionalized phase introduces a finite density of anyons, whose subsequent condensation leads to daughter states. This correspondence can be made more precise via the generalized stack-and-condense framework developed in \cite{seo2026unified}. Building on the hierarchy-wavefunction formalism for encoding anyon condensation \cite{haldane1983fractional,halperin1984statistics}, we construct explicit superconducting wavefunctions obtained by the condensation of charged anyons.

Our main results are the followings. First, we construct wavefunctions for several paradigmatic anyon superconductors, including those descending from the semion {state}, a $\nu=2/3$ hierarchy state, a $\nu=1/3$ Laughlin state, and Pfaffian state. We establish superconductivity through off-diagonal long-range orders (ODLRO) in wavefunctions and determine condensate charge, chiral central charge, and residual topological order from the associated topological field theories. Notably, this construction makes their non-{Bardeen-Cooper-Schrieffer (BCS) origin manifest: superconductivity arises from anyon condensation rather than Cooper pairing of electrons. Second, for the semion case, using quasilocal operator formalism~\cite{hansson2007conformal,hansson2007composite-fermion,hansson2009conformal,hansson2009quantum,suorsa2011general,suorsa2011quasihole,hansson2017quantum}, we show that our superconducting wavefunction of many semions is equivalent to Laughlin's original ansatz \cite{laughlin1988superconducting,laughlin1988relationship}. Third, extending beyond anyons, we apply the construction to electrons in a $\nu=1$ integer quantum Hall state and ``condense" them, obtaining wavefunctions for charge-$2e$ non-Abelian chiral superconductors. Finally, we connect our construction to the anyon-Hilbert-space formalism \cite{yan2026anyon,li2026bound}, showing how these wavefunctions can naturally emerge from multi-anyon states in the dilute limit. Together, these results forge a unified understandings of anyon condensation, wavefunctions, and field theory.
 
\textbf{\color{brown}{1.~General Framework:}} Our formalism bases on the quantum Hall hierarchy construction \cite{haldane1983fractional,halperin1984statistics}. The hierarchy wavefunction $\Psi_{1}$, obtained by condensing anyons of a parent topological phase $\Psi_0$, takes the form \cite{haldane1983fractional,halperin1984statistics}
\begin{align}
    \Psi_{1}(\{z_i\})
    =
    \int {\prod_a d^2 u_a}\,
    \Phi^{*}_1(\{u_a\})\,
    \Psi_{0}(\{z_i\};\{u_a\}).
    \label{eq:hierarchy_wavefunction}
\end{align}
Here, $z_i$ and $u_a$ denote the holomorphic coordinates of the electrons and the anyons, respectively, and $\Psi_{0}(\{z_i\};\{u_a\})$ is the parent-state wavefunction with anyons inserted at $u_a$. The pseudowavefunction $\Phi(\{u_a\})$ specifies the correlated many-anyon state into which the anyons condense. Each component of Eq.~\eqref{eq:hierarchy_wavefunction} is a correlation function of conformal field theory \cite{moore1991nonabelions,bonderson2008fractional} 
\begin{align}
    \Psi_0(\{z_i\}; \{u_a\}) &= \left\langle \prod_{i = 1}^{N_e} V_e(z_i) \prod_{a = 1}^{N_q} V_q(u_a) \right\rangle, \nonumber \\
    \Phi_1^{*}(\{u_a\}) &= \left\langle \prod_{a = 1}^{N_q} \bar{V}_{q'}(u_a) \right\rangle, \nonumber
\end{align}
where $V_e$ are the fermionic electron operators, while $V_q$ and $\bar{V}_{q'}$ are the anyonic quasiparticle operators. For the hierarchy wavefunction to be well defined, the product operator $V_q \bar{V}_{q'}$ must be bosonic. Here we suppress the neutralizing background charge to keep the expressions simple. This formalism Eq.~\eqref{eq:hierarchy_wavefunction} explicitly identifies the anyons that drives a transition from the parent state $\Psi_0$ to the daughter state $\Psi_1$ upon condensation.

The plasma analogy maps the wavefunction probability density, $|\Psi|^2\propto e^{-\mathcal H}$, to an effective Hamiltonian $\mathcal H$ in terms of the species-resolved density vector $\vec{\rho}(\mathbf r)$ \cite{laughlin1983anomalous,degail2008plasma}:
\begin{equation}
    \mathcal H
    =
    -\frac{1}{2}\int_{r, r'}
    \vec{\rho}(\mathbf r)^{\mathsf T}
    M_K
    \vec{\rho}(\mathbf r')
    \ln {|\mathbf r-\mathbf r'|} 
    +
    \int_{r}
    \vec{\rho}(\mathbf r)^{\mathsf T}
    \vec{t}_p|\mathbf r|^2. 
    \nonumber
\end{equation}
Here, $M_K$ and $\vec t_p$ denote the exponent matrix and plasma charge vector. The quantum Hall wave function is stable and incompressible when $M_K$ is positive definite \cite{degail2008plasma}. Then, the superconductivity is signaled by a positive-semidefinite $M_K$ with null vector $\vec v_0$, whose associated operator $\mathcal O_{\mathrm{SC}}$ exhibits ODLRO. The ODLRO is diagnosed by the reduced density matrix \cite{yang1962concept}
\begin{equation}
    \rho(\eta, \eta') = \frac{\int_z \Psi_1^*(\eta; \{z_i\}) \Psi_1(\eta'; \{z_i\})}{\int_z \Psi_1^*(\{z_i\}) \Psi_1(\{z_i\})}, \nonumber
\end{equation}
where 
\begin{align*}
    &\Psi_{1}(\eta; \{z_i\}) \\
    &= \int_u \left\langle {\mathcal {O_{\mathrm{SC}}}(\eta)} \prod_{i = 1}^{N_e} V_e(z_i) \prod_{a = 1}^{N_q} V_q(u_a){\prod_{a = 1}^{N_q} \bar{V}_{q'}(u_a)} \right\rangle,
\end{align*} 
denotes the hierarchy wavefunction with {$\mathcal O_{\mathrm{SC}}$} inserted at $\eta$. The ODLRO corresponds to  
\begin{equation} 
    \lim_{\abs{\eta - \eta'} \to \infty} \rho(\eta, \eta') = \mathrm{const.} \neq 0. \nonumber
\end{equation}
Here, $\mathcal O_{\mathrm{SC}}$ carries electric charge $q_0=\vec v_0^{\,\mathsf T}\vec t$, where $\vec t$ is the physical electric charge vector, {which is in general distinct from} $\vec t_p$. Thus, the ODLRO signals the charge-$q_0$ superconductivity. Hierarchy formalism and plasma analogy are reviewed in \cite{supp}.

For each hierarchy wavefunction in Eq.~\eqref{eq:hierarchy_wavefunction}, we can identify the associated topological field theory \cite{supp}, from which the condensate charge, chiral central charge, and residual topological order are read off. For example, for Abelian states with $K$-matrix, we prove that the plasma null vector satisfies $K\vec v_0=0$, yielding \cite{supp}
\begin{equation}
    \mathcal L 
    =
    \frac{q_0}{2\pi}A\,d\alpha
    +\frac{1}{2g^2}(d\alpha)^2+\cdots, \nonumber
    \label{eq:eff_qft_sc}
\end{equation}
which describes a charge-$q_0$ superconductor. The dictionary between wavefunctions and field theory is in \cite{supp}. 
 
\textbf{\color{brown}{2.~Laughlin's semion case:}} We first consider Laughlin's semion superconductor \cite{laughlin1988superconducting}, where doped charge-$e/2$ semions condense to produce a charge-$e$ superconductor. We begin with the conformal-block wavefunction of bosonic $\nu=1/2$ state with the semions 
\begin{align}
    \Psi_0 
    =
    \left\langle
    \prod_{\alpha = 1}^{N_1}
    e^{i \phi_1(z_\alpha)}
    \prod_{\beta = 1}^{N_2}
    e^{i \phi_2(u_\beta)}
    \right\rangle. 
    \label{eq:semion_parent} 
\end{align}
The pseudowavefunction of the semions is 
\begin{align}
    \Phi_1^{*} 
    =
    \int_{w}
    \left\langle
    \prod_{\alpha = 1}^{N_2}
    e^{- i \bar{\phi}_2(\bar{u}_\alpha)}
    \prod_{\beta = 1}^{N_3}
    e^{- i \bar{\phi}_3(\bar{w}_\beta)}
    \right\rangle,
    \label{eq:semion_pseudowavefunction}
\end{align}
where the $w_\beta$ are auxiliary hierarchy coordinates that are integrated over. The chiral and antichiral bosons satisfy 
$\langle \phi_I (z) \phi_J (w) \rangle = - \kappa_{IJ} \log(z - w)$, and $\langle \bar{\phi}_I (z) \bar{\phi}_J (w) \rangle = - \bar{\kappa}_{IJ} \log (\bar{z} - \bar{w})$. 
where 
\begin{equation}
    \kappa = 
    \begin{pmatrix}
        2 & -1 & 0 \\
        -1 & \frac{1}{2} & 0 \\
        0 & 0 & 0
    \end{pmatrix}, 
    \quad
    \bar{\kappa} = 
    \begin{pmatrix}
        0 & 0 & 0 \\
        0 & \frac{1}{2} & -1 \\
        0 & -1 & 2
    \end{pmatrix}.
    \nonumber
\end{equation}
The conjugated $w$ sector realizes the antisemion order, required to produce superconductivity in the generalized stack-and-condense framework~\cite{seo2026unified}. The result is \cite{supp} 
\begin{equation}
    \Psi_1
    = (z-z)^2 \int_u (z-u)^{-1} |u-u| \int_w (\bar{u}-\bar{w})^{-1} (\bar{w}-\bar{w})^2. 
    \nonumber
\end{equation} 
Products over coordinate differences and integral measures are abbreviated for simplicity, e.g., $(z-z)\equiv\prod_{\alpha<\beta}(z_\alpha-z_\beta)$. Gaussian factors are suppressed as well. We note that the wavefunction only describes the long-range behavior and terms like $(z - u)^{-1}$ is expected to be replaced by a smooth function when $z \to u$.

Its exponent matrix in the $(z,u,w)$ basis is
\begin{equation}
    M_K = 
    \begin{pmatrix}
        4 & -2 & 0 \\
        -2 & 2 & -2 \\
        0 & -2 & 4
    \end{pmatrix}, \nonumber 
\end{equation}
which has a null vector $\vec{v}_0^\mathsf{T} =(1, 2, 1)$ and the charge vector $\vec{t}^\mathsf{T} =(1, 0, 0)$ \cite{supp}. Consequently, the following operator with $q_0 =\vec{t}^\mathsf{T} \vec{v}_0 = 1$ develops ODLRO \cite{supp} 
\begin{align}
    \mathcal{O}_{\text{SC}} (\eta) \sim \exp{i\vec{v}_0^\mathsf{T} \vec{\varphi}(\eta ,\bar{\eta})} = e^{i [\varphi_1(\eta) + 2 \varphi_2(\eta, \bar{\eta}) + \varphi_3(\bar{\eta})]}, \nonumber 
\end{align}
where $\vec{\varphi} = (\varphi_1, \varphi_2, \varphi_3)^\mathsf{T}$ is defined in terms of $\phi_i$ and $\bar{\phi}_i$ in Eqs.~\eqref{eq:semion_parent} and \eqref{eq:semion_pseudowavefunction}:
$\varphi_1(\eta) = \phi_1(\eta), \quad  \varphi_2(\eta, \bar{\eta}) = \phi_2(\eta) - \bar{\phi}_2(\bar{\eta})$, and $\varphi_3 (\bar{\eta}) = -\bar{\phi}_3 (\bar{\eta})$. This establishes the emergence of charge-$e$ superfluidity.

Using the wavefunction--field-theory dictionary \cite{supp}, e.g.,
$(z-z)^2\to-\frac{2}{4\pi}a\,da$, we read the field theory
\begin{equation}
    K=
    \begin{pmatrix}
        2 & -1 & 0 \\
        -1 & 0 & 1 \\
        0 & 1 & -2
    \end{pmatrix},
    \qquad
    \vec t=
    \begin{pmatrix}
        1 \\ 0 \\ 0
    \end{pmatrix}.
    \nonumber
\end{equation}
As expected, the null vector $\vec v_0$ of $M_K$ also satisfies 
$K\vec v_0=0$. An $\mathrm{SL}(3,\mathbb Z)$ transformation gives \cite{supp}
\begin{equation}
    K'=
    \begin{pmatrix}
        0 & -1 & 0 \\
        -1 & 2 & 0 \\
        0 & 0 & 0
    \end{pmatrix},
    \qquad
    \vec t'=
    \begin{pmatrix}
        0 \\ 1 \\ 1
    \end{pmatrix}.
    \nonumber
\end{equation}
This isolates the charge-$e$ superfluid mode in the final component, while the $2\times2$ block is topologically trivial, i.e., the effective theory is equivalent to Eq.~\eqref{eq:eff_qft_sc}. Finally, the absence of topological order and the vanishing chiral central charge agree with previous analyses~\cite{seo2026unified, shi2025doping}.

Remarkably, the pseudowavefunction Eq.~\eqref{eq:semion_pseudowavefunction} can be recast into Laughlin's original ansatz \cite{laughlin1988superconducting}. By rewriting the hierarchy integral in terms of quasilocal operators~\cite{suorsa2011general}, we can prove~\cite{supp}
\begin{align}
    \Phi_1 \propto \prod_{\alpha < \beta} \left(\frac{u_\alpha - u_\beta}{\abs{u_\alpha - u_\beta}}\right)^{\frac{1}{2}}~ \cdot ~ \text{det} ~\psi_{\text{CF}}^{\text{2LL}}\left(\{u_i\}_{i=1}^{N_1}\right), \nonumber \label{eq:semion_quasilocal}
\end{align}
where $\psi_{\text{CF}}^{\text{2LL}}\left(\{u_i\}_{i=1}^{N_1}\right)$ describes fermionized semions filling two Landau levels, as in \cite{laughlin1988superconducting}. This provides a nontrivial bridge between field theory, category framework, hierarchy construction, and Laughlin's ansatz \cite{laughlin1988superconducting}.

The construction extends straightforwardly to a $\nu=2/3$ state doped with charge-$2e/3$ anyons. The resulting wavefunction exhibits charge-$2e$ ODLRO, while the corresponding field theory yields a chiral central charge $c=-2$ and no residual topological order \cite{supp}, consistent with previous analyses \cite{shi2025doping,shi2025dopinglattice}. Furthermore, the pseudowavefunction can be recast as a determinant of the fermionized anyons occupying the three Landau levels, as expected \cite{laughlin1988superconducting}.

\textbf{\color{brown}{3.~Doping $e/3$-anyons to Laughlin state:}} Non-Abelian chiral superconductors follow from the same construction. Here we consider doping the $\nu=1/3$ Laughlin state with charge-$e/3$ anyons. The parent state is 
\begin{equation}
    \Psi_0 = \left\langle \prod_\alpha e^{i \phi_1(z_\alpha)} \prod_\alpha e^{i \phi_2(u_\alpha)} \right\rangle, \nonumber 
\end{equation}
and the pseudowavefunction is 
\begin{equation}
    \Phi_1^{*} = \int_{w} \left\langle \prod_\alpha e^{- i \bar{\phi}_2(\bar{u}_\alpha)} \prod_\alpha \psi(\bar{w}_\alpha) e^{- i \bar{\phi}_3(\bar{w}_\alpha)} \right\rangle. \nonumber 
    \label{eq:sc_laughlin_pseudo}
\end{equation}
Here, we use {the chiral and antichiral bosons with 
\begin{equation}
    \kappa = 
    \begin{pmatrix}
        3 & -1 & 0 \\
        -1 & \frac{1}{3} & 0 \\
        0 & 0 & 0
    \end{pmatrix},
    \quad
    \bar{\kappa} = 
    \begin{pmatrix}
        0 & 0 & 0 \\
        0 & \frac{1}{3} & -1 \\
        0 & -1 & 3 
    \end{pmatrix}.
    \nonumber
\end{equation}
The resulting wavefunction is then~\cite{supp}
\begin{align}
    \Psi_1 = (z - z)^3 &\int_u (z - u)^{-1} \abs{u - u}^{\frac{2}{3}} \nonumber \\ 
    \times &\int_w (\bar{u} - \bar{w})^{-1} (\bar{w} - \bar{w})^3 \, \operatorname{Pf}\!\left(\frac{1}{\bar{w} - \bar{w}}\right). 
    \label{eq:sc_laughlin}
\end{align}
Using the the plasma analogy of the Pfaffian \cite{bonderson2011plasma}
\begin{equation}
    \abs{\operatorname{Pf}\!\left(\frac{1}{z - z}\right)}^2 = \int_\zeta \abs{\zeta - \zeta}^3 \abs{\zeta - z}^{-3} \abs{z - z}^3, \nonumber 
\end{equation}
we find its exponenent matrix~\cite{supp}
\begin{equation}
    M_K=
    \begin{pmatrix}
        6 & -2 & 0 & 0 \\
        -2 & \frac{4}{3} & -2 & 0 \\
        0 & -2 & 9 & -3 \\
        0 & 0 & -3 & 3
    \end{pmatrix},
    \nonumber
\end{equation}
which is positive semidefinite. The algebraic kernel of $M_K$ is spanned by $(1,3,1,1)^{\mathsf T}$. However, because the Pfaffian sector requires an even number of electrons, physical states lie in a fixed fermion-parity sector; hence the smallest admissible null vector is $\vec v_0=(2,6,2,2)^{\mathsf T}$~\cite{read2000paired}. The associated vertex operator is
\begin{equation}
    \mathcal O_{\mathrm{SC}}(\eta)
    = e^{i [2 \phi_1(\eta) + 6 (\phi_2(\eta) - \bar{\phi}_2 (\bar{\eta})) - 2 \bar{\phi}_3(\bar{\eta})]},
    \nonumber
\end{equation}
which exhibits ODLRO. Its electric charge $q_0=2$ identifies the state as a charge-$2e$ superconductor.

We next derive the field theory using the dictionary \cite{supp}. First, the factor $(z-z)^3$ in Eq.~\eqref{eq:sc_laughlin} corresponds to 
\begin{align}
    \mathcal L
    =
    -\frac{3}{4\pi}a\,da
    +\frac{1}{2\pi}A\,da
    +\mathcal L_{\mathrm{aux}}[a], \nonumber 
\end{align}
where $\mathcal L_{\mathrm{aux}}$ encodes the $u$ and $w$ sectors. The factors $(z-u)^{-1}$ and $(\bar u-\bar w)^{-1}$ then contribute
\begin{align}
    \mathcal L_{\mathrm{aux}}[a]
    =
    \frac{1}{2\pi}a\,d\hat{b}
    -\frac{1}{2\pi}\hat{b}\,d\beta
    +\cdots ,
    \nonumber
\end{align}
where $\hat{b}$ and $\beta$ are $\mathrm{U}(1)$ gauge fields. Finally, the remaining antiholomorphic factors with Pfaffian in $w$ result in 
\begin{align}
    &\frac{2}{4 \pi} \Tr(c d c + \frac{2}{3} c^3) - \frac{2}{4 \pi} \Tr(c) d \Tr(c) \nonumber \\
	&+ \frac{2}{4 \pi} \beta d \beta + \frac{1}{2 \pi} \beta d \Tr(c),
    \nonumber
\end{align}
where $c$ is a $\mathrm{U}(2)$ gauge field. Integrating out $\hat{b}$ and $\beta$ in order, we find
\begin{align}
    \mathcal L &= \frac{2}{4 \pi} \Tr(c d c + \frac{2}{3} c^3) - \frac{1}{4 \pi} \Tr(c) d \Tr(c) \nonumber \\
    &\quad + \frac{1}{2 \pi} A d \Tr(c) + \mathrm{CS}[A, g], \nonumber
\end{align}
which describes a charge-$2e$ non-Abelian chiral superconductor with $c=-1/2$, in agreement with \cite{shi2025nonabelian}. Choosing the opposite chirality for the Ising sector in Eq.~\eqref{eq:sc_laughlin_pseudo} yields another superconductor with $c = 1 / 2$}, as detailed in \cite{supp}.

\textbf{\color{brown}{4.~Doping semions to Pfaffian:}} Similarly, we consider a Pfaffian state doped with semions. The derivation closely parallels that for the non-Abelian superconductor above and is deferred to~\cite{supp}. The parent state is 
\begin{equation}
    \Psi_0
    =
    \left\langle
    \prod_{\alpha=1}^{N_1}
    \psi(z_\alpha)e^{i \phi_1(z_\alpha)}
    \prod_{\beta=1}^{N_2}
    e^{i \phi_2(u_\beta)}
    \right\rangle, \nonumber
\end{equation}
where we use the same chiral bosons and pseudowavefunction Eq.~\eqref{eq:semion_pseudowavefunction} as the semion case. The result is
\begin{align}
    \Psi_1 = \operatorname{Pf}\!\left(\frac{1}{z - z}\right) &(z - z)^2 \int_u (z - u)^{-1} \abs{u - u} \nonumber\\
    &\times \int_w (\bar{u} - \bar{w})^{-1} (\bar{w} - \bar{w})^2, \nonumber 
\end{align}
whose exponent matrix has the null vector with $q_0 = 2$, thereby establishing charge-$2e$ superconductivity. To further characterize the state, we derive its effective field theory using the dictionary~\cite{supp}, which identifies it as a charge-$2e$ superconductor with $c=1/2$ and no residual topological order, consistent with~\cite{seo2026unified,shi2025dopinglattice}.

\textbf{\color{brown}{5.~Electron ``condensate'' in $\nu=1$ state:}} Having consistently reproduced known anyon superconductors \cite{laughlin1988superconducting,shi2025doping,shi2025dopinglattice,shi2025nonabelian}, we now extend the construction to electrons in a $\nu=1$ integer quantum Hall state, treating formally fermions as anyonic excitations and condensing them. The details of the derivation is in~\cite{supp}.

The parent state is 
\begin{equation}
    \Psi_0 = \left\langle \prod_{\alpha = 1}^{N_1} e^{i \phi_1(z_\alpha)} \prod_{\beta = 1}^{N_2} e^{i \phi_2(u_\beta)} \right\rangle, \nonumber 
\end{equation}
where $\langle \phi_I(z) \phi_J(w) \rangle = - \kappa_{IJ} \log(z - w)$ with
\begin{equation}
    \kappa = 
    \begin{pmatrix}
        1 & -1 \\
        -1 & 1
    \end{pmatrix}. \nonumber 
\end{equation}
We choose the pseudowavefunction as 
\begin{equation} \label{eq:IQH_pseudo}
    \Phi_1^{*} = \left\langle \prod_{\alpha = 1}^{N_2} \bar{\psi}(u_\alpha) \right\rangle, 
\end{equation}
then the resulting wavefunction is 
\begin{equation}
    \Psi_1 = (z - z) \int_u (z - u)^{-1} (u - u) \operatorname{Pf}\!\left(\frac{1}{\bar{u} - \bar{u}}\right). \nonumber 
\end{equation}
The corresponding exponent matrix has a null vector whose associated 
operator carries charge $q_0=2$, establishing charge-$2e$ 
superconductivity~\cite{supp}. 

The effective field theory derived from the wavefunction also identifies the state as a charge-$2e$ superconductor with $c=1/2$ and no residual topological order~\cite{supp}. Intriguingly, this phase resembles the 
$p+ip$ superconductor obtained by proximity coupling a $\nu=1$ integer 
quantum Hall state to an $s$-wave superconductor~\cite{qi2010chiral}. Choosing the opposite chirality for the Ising sector in Eq.~\eqref{eq:IQH_pseudo} yields another superconductor with $c=3/2$~\cite{supp}. 
 
\textbf{\color{brown}{6.~Anyon Hilbert Space Formalism:}} We show that the pseudowavefunctions Eq.~\eqref{eq:hierarchy_wavefunction} can naturally emerge from the anyon-Hilbert-space formulation~\cite{li2026bound}, which applies to ideal Chern bands and moir\'e bands of twisted bilayer $\mathrm{MoTe}_2$ \cite{yan2026anyon}. Specifically, they emerge as the long-distance limit of multi-anyon states when inter-anyon separations greatly exceed the effective magnetic length.

In the formulation of~\cite{li2026bound}, an $N_h$-quasihole state$\ket{\xi}\equiv\ket{\xi_1,\ldots,\xi_{N_h}}$ of the $\nu=1/q$ Laughlin state satisfies
\begin{equation}
    \braket{z_1,\ldots,z_{N_e}}{\xi}
    \propto
    \prod_{i=1}^{N_e}\prod_{j=1}^{N_h}(z_i-\xi_j)
    \prod_{i<j}^{N_e}(z_i-z_j)^q.
    \nonumber
\end{equation}
Since the states $\{\ket{\xi}\}$ form an overcomplete basis, a general state $\ket f$ is represented by the antiholomorphic function $f(\bar\xi)\equiv\braket{\xi}{f}$. Defining the K\"ahler potential by $e^{\mathcal K(\bar\omega,\xi)}=\braket{\omega}{\xi}$, the measure $d\mu_\xi$ is fixed as 
$\int d\mu_\xi\, e^{\mathcal K(\bar\omega,\xi)}f(\bar\xi)=
f(\bar\omega)$. Equivalently, we can write the identity resolution as 
\begin{equation}
    \int d\mu_\xi\,\ket{\xi}\bra{\xi}
    =
    \mathds{1}_{\mathcal H_{\rm anyon}}.
    \nonumber
\end{equation}

We now consider the dilute anyon gas regime in which all inter-quasi-hole separations greatly exceed the magnetic length. In this limit, the quasihole basis $\{\ket{\xi}\}$ becomes an orthogonal basis since now quasiholes at different coordinates have a vanishing overlap. Thus,
\begin{align}
e^{\mathcal{K}(\bar{\omega}, \xi)} = \braket{\omega}{\xi} = e^{\mathcal{K}_\xi} \delta_{\omega, \xi}. \nonumber 
\end{align} 
Consequently, the identity resolution is 
\begin{equation}
    \mathds{1} = \int d^2 \xi ~ e^{- \mathcal{K}_\xi} \ket{\xi} \bra{\xi}, \nonumber
\end{equation}
implying $d \mu_\xi = d^2 \xi \, e^{- \mathcal{K}_\xi}$. Using the plasma analogy \cite{supp}, we can obtain 
\begin{equation}
    e^{- \mathcal{K}_\xi} = \exp{\frac{2}{q} \sum_{i < j}^{N_h} \ln\abs{\xi_i - \xi_j}}. \nonumber
\end{equation}
Plugging this in, we finally find 
\begin{equation}
    f(z) \equiv \braket{z}{f} = (z - z)^q \int_\xi (z - \xi) \abs{\xi - \xi}^{\frac{2}{q}} f(\bar{\xi}). \nonumber
\end{equation}
This is precisely a hierarchy pseudowavefunction
\begin{equation}
    \Phi_1(\{\xi\}) = (\xi - \xi)^{\frac{1}{q}} f(\xi). \nonumber
\end{equation}

The connection between the hierarchy construction and the anyon-Hilbert-space formulation can be understood intuitively from their common long-distance limits. Our derivation applies when all inter-anyon separations well exceed the magnetic length $\ell_B$. In this regime, short-distance microscopic details become irrelevant, while the universal topological data remain. Hence, the anyon-Hilbert-space formulation in this limit reduces to the form entering the hierarchy integral, which is designed to capture precisely this universal long-distance structure. {In this sense, the divergent terms in our wavefunctions like $(z - u)^{-1}$ can be naturally viewed as the long-wavelength behavior when $|z-u| \gg \ell_B$. In the short distance regime, the divergent terms are expected to be regularized as a smooth function. Similar arguments have also appeared in quantum Hall literature \cite{read1990excitation}.}
 
\textbf{\color{brown}{7.~Conclusions:}} We developed a hierarchy-wavefunction framework for superconducting states obtained by condensing charged anyons of a parent topological order. Null vectors of the plasma exponent matrix diagnose charged ODLRO, while the corresponding effective field theories determine chiral central charge and residual topological order. We applied the construction to the semion {state}, a $\nu=2/3$ hierarchy state, {the} $\nu=1/3$ Laughlin state, and the Pfaffian state, obtaining Abelian and non-Abelian superconductors. In particular, for the semion case, we related the pseudowavefunctions to Laughlin's original ansatz, i.e., fermionized anyons filling two effective Landau levels. Extending the same logic to electrons in a $\nu=1$ integer quantum Hall state yielded explicit wavefunctions for charge-$2e$ non-Abelian chiral superconductors. We also connected hierarchy integrals to the dilute, long-distance limit of the anyon-Hilbert-space formulation, which is directly relevant to twisted bilayer MoTe\textsubscript{2}. By unifying many-body wavefunctions with anyon condensation and topological field theory, our framework provides a concrete foundation for exploring anyon superconductivity beyond the BCS paradigm. 

\begin{acknowledgments}
    We thank Eslam Khalaf, Yong Baek Kim, Zi-Yang Meng, and Jie Wang for helpful discussions. This work is financially supported by Samsung Science and Technology Foundation under Project Number SSTF-BA2401-03, the NRF of Korea (Grants No. RS-2026-25479545, RS-2024-00410027, RS-2024-00444725, RS-2023-00256050, RS-2025-25453111, RS-2025-08542968) funded by the Korean Government (MSIT), the Air Force Office of Scientific Research under Award No. FA23862514026, and Institute of Basic Science under project code IBS-R014-D1. T.~L. is partially supported by KAIST Undergraduate Research Program (URP).
\end{acknowledgments}

\bibliography{ref}

\onecolumngrid

\clearpage

\begin{center}
    \textbf{\large Supplemental Material of ``Wavefunctions for Anyon Superconductors''}
\end{center}

\section{Review of hierarchy construction and conformal field theory techniques}

In this section, we review the framework of quantum Hall hierarchy construction \cite{haldane1983fractional,halperin1984statistics} and the use of conformal field theory correlation functions to build quantum Hall wavefunctions \cite{moore1991nonabelions,bonderson2008fractional}.

Quantum Hall wavefunctions can be constructed as correlation functions of conformal field theory \cite{moore1991nonabelions}. For example, the Laughlin state at filling fraction $\nu = 1 / m$ is given by 
\begin{equation}
    \Psi(z_1, \dots, z_N) = \left\langle \prod_{i = 1}^N e^{i \sqrt{m} \varphi(z_i)} \, \mathcal{O}_\mathrm{bg} \right\rangle = \prod_{i < j}^N (z_i - z_j)^m \, e^{- \frac{1}{4} \sum_i \abs{z_i}^2},
\end{equation}
where $\varphi(z)$ is a chiral boson field obeying $\langle \varphi(z) \varphi(w) \rangle = - \log(z - w)$ and $\mathcal{O}_\mathrm{bg}$ is the background neutralizing charge. We have kept the normal ordering of the vertex operator implicit and set the magnetic length to $1$ for notational simplicity. The non-Abelian Moore-Read state can also be constructed by using the Ising conformal field theory, as 
\begin{equation}
    \Psi(z_1, \dots, z_N) = \left\langle \prod_{i = 1}^N \psi(z_i) e^{i \sqrt{m} \varphi(z_i)} \, \mathcal{O}_\mathrm{bg} \right\rangle = \operatorname{Pf}\!\left(\frac{1}{z_i - z_j}\right) \prod_{i < j}^N (z_i - z_j)^m \, e^{- \frac{1}{4} \sum_i \abs{z_i}^2}.
\end{equation}
Here, $\psi$ is the fermionic field in the Ising conformal field theory.

Given a quantum Hall wavefunction, one can construct a wavefunction at a different filling fraction via the hierarchy construction \cite{haldane1983fractional,halperin1984statistics}. The underlying physical picture is that anyons emerge due to the change of the filling fraction and condense to their own incompressible state. Following this, given a parent quantum Hall wavefunction $\Psi_0$, the first-level hierarhcy state is 
\begin{equation}
    \Psi_1(z_1, \dots, z_N) = \int \prod_{a = 1}^M d^2 u_a \, \Phi^*(u_1, \dots, u_M) \Psi_0(z_1, \dots, z_N; u_1, \dots, u_M),
\end{equation}
where $u_a$ are the coordinates of the emergent anyons. Here, $\Phi^*$ is called the pseudowavefunction and describes the many-body state that the quasiparticles condense to. The pseudowavefunctions can also be constructed as correlators in conformal field theory.

While the canonical basis where chiral bosons obey $\langle \varphi_I(z) \varphi_J(w) \rangle = - \delta_{IJ} \log(z - w)$ is widely used in literature, there exists the one which we will refer to as the $K$ matrix basis \cite{suorsa2011general} where the topological properties of the wavefunction is more explicit. In the $K$-matrix basis, we have
\begin{align}
    \langle \phi_I(z) \phi_J(w) \rangle = - \kappa_{IJ} \log(z - w), \\ 
    \langle \bar{\phi}_I(z) \bar{\phi}_J(w) \rangle = - \bar{\kappa}_{IJ} \log(\bar{z} - \bar{w}),
\end{align}
where $\kappa$ and $\bar{\kappa}$ are symmetric positive-semidefinite satisfying $K = \kappa - \bar{\kappa}$. The choice of $\kappa$ and $\bar{\kappa}$ is not unique. To map one basis to the other, we decompose $K$ as 
\begin{equation}
    K = O D O^\mathsf{T},
\end{equation}
where $O$ is an orthogonal matrix and $D$ is a diagonal matrix, which is always possible since $K$ is a symmetric matrix. Write 
\begin{equation}
    D = \sqrt{\abs{D}} \eta \sqrt{\abs{D}}
\end{equation}
where $\eta$ is the signature matrix, and define 
\begin{equation}
    E = O \sqrt{\abs{D}}.
\end{equation}
Then, the $K$-matrix chiral bosons are given by 
\begin{equation} \label{eq:Kmatrix_basis_transf}
    \phi_I = \sum_a E_{Ia} \varphi_a. 
\end{equation}

\section{Review of plasma analogy}

The probability density of quantum Hall wavefunctions can be mapped to as two-dimensional plasma \cite{laughlin1983anomalous}. For many Abelian quantum Hall states, we have 
\begin{equation}
    \abs{\Psi}^2 = \int_u \exp{- \mathcal{H}},
\end{equation}
where \cite{degail2008plasma}
\begin{equation}
    \mathcal{H}[\vec{\rho}(r)] = - \frac{1}{2} \iint d^2 r d^2 r' \, \vec{\rho}(r)^\mathsf{T} M_K \vec{\rho}(r') \ln\abs{r - r'} + \int d^2 r \, \vec{\rho}(r)^\mathsf{T} \vec{t}_p \abs{r}^2.
\end{equation}
Here, $\vec{\rho}(r)$ is the vector of (quasi)particle densities and the integral is over the quasiparticle coordinates. We will refer to $M_K$ and $\vec{t}_p$ as the exponent matrix and plasma charge vector, respectively. The plasma analogy has been partially generalized to non-Abelian quantum Hall states \cite{gurarie1997plasma,bonderson2011plasma}, yielding the plasma analogy of the Pfaffian \cite{bonderson2011plasma}
\begin{equation}
    \abs{\operatorname{Pf}\!\left(\frac{1}{z - z}\right)}^2 = \int_w (z - z)^3 (z - w)^{-3} (w - w)^3.
\end{equation}

The plasma may be unstable and thus the corresponding quantum Hall wavefunction is unphysical, depending on $M_K$ and $\vec{t}_p$ \cite{degail2008plasma}. Note that the effective energy functional of the plasma $\mathcal{H}$ admits a minimal-energy configuration if and only if $M_K$ is positive-semidefinite. Then, the minimal-energy configuration is given as the solution of \cite{degail2008plasma}
\begin{equation}
    M_K \vec{\rho}(r) = \vec{t}_p.
\end{equation}
In particular, if $M_K$ is positive-definite, then the plasma has a unique minimal-energy configuration and takes finite amount of energy to fluctuate the densities. Consequently, the associated quantum Hall wavefunction becomes an incompressible liquid. On the other hand, if $M_K$ has a zero eigenvalue, then the ground-state subspace is multidimensional given that the null vector of $M_K$ is orthogonal to $\vec{t}_p$. (We will show in a later section that this condition is automatically satisfied if we use the conformal field theory techniques.) The density fluctuation within the subspace does not cost energy and the corresponding wavefunction becomes compressible \cite{degail2008plasma}. If the gapless mode carries a nontrivial charge, the resulting state is identified as a superconductor.

\section{Wavefunction--Chern-Simons theory dictionary}

In this section, we outline how the effective Chern-Simons theory is written down when an anyon-superconduncting wavefunction is given. We assume that the wavefunction is given by a product of $(z - z)^m$, $(z - u)^n$, $\operatorname{Pf}[1/(u - u)] (u - u)^M$, or their complex conjugates.

Consider a term of the form $(z - z)^m$. Since it describes the Laughlin state with filling fraction $\nu = 1 / m$, it contributes to the Chern-Simons Lagrangian with the term $(- m / 4 \pi) a d a$, where $a$ is a dynamical $\mathrm{U}(1)$ gauge field. If the term is complex-conjugated, i.e., $(\bar{z} - \bar{z})^m$, then it contributes by $(m / 4 \pi) a d a$. In the same vein, a term of the form $(z - u)^n$ dictates mutual braiding statistics between different kinds of particles, it contributes by $(- n / 2 \pi) a d b$, where $b$ is another dynamical $\mathrm{U}(1)$ gauge field. Similarly, if the term is complex-conjugated, the sign of the mutual Chern-Simons level is reversed. Lastly, the term $\operatorname{Pf}[1 / (u - u)] (u - u)^M$ is described by 
\begin{equation}
    \mathcal{L} = - \frac{2}{4 \pi} \Tr(c d c + \frac{2}{3} c^3) + \frac{2}{4 \pi} \Tr(c) d \Tr(c) - \frac{M - 1}{4 \pi} \beta d \beta + \frac{1}{2 \pi} \beta d (A - \Tr(c)),
\end{equation}
where $\beta$ and $c$ are $\mathrm{U}(1)$ and $\mathrm{U}(2)$ gauge fields and $A$ is the background field. For the term $\operatorname{Pf}[1 / (u - u)] (\bar{u} - \bar{u})^M$, the corresponding Chern-Simons theory is 
\begin{equation}
    \mathcal{L} = - \frac{2}{4 \pi} \Tr(c d c + \frac{2}{3} c^3) + \frac{2}{4 \pi} \Tr(c) d \Tr(c) + \frac{M + 1}{4 \pi} \beta d \beta + \frac{1}{2 \pi} \beta d (A - \Tr(c)).
\end{equation}
For $M = 0$, the Lagrangian reduces to 
\begin{equation}
    \mathcal{L}_{p + ip} = - \frac{2}{4 \pi} \Tr(c d c + \frac{2}{3} c^3) + \frac{1}{4 \pi} \Tr(c) d \Tr(c) + \frac{1}{2 \pi} \Tr(c) d A - \mathrm{CS}[A, g]
\end{equation}
which describes a $p + ip$ superconductor, as expected \cite{read2000paired}.

For completeness, we here present a derivation of the effective field theory describing $\operatorname{Pf}[1 / (u - u)] (\bar{u} - \bar{u})^M$. The derivation for $\operatorname{Pf}[1 / (u - u)] (u - u)^M$ can be found in Ref.~\cite{shi2025dopinglattice}. The filling fraction of $\operatorname{Pf}[1 / (u - u)] (\bar{u} - \bar{u})^M$ is given by $\nu = - 1 / M$, and the state is fermionic or bosonic when $M$ is even or odd, respectively. We will assume that $M$ is odd and the state is bosonic. The generalization to the fermionic state should be straightforward.

We first decompose the microscopic bosonic operator into two fermionic operators, as $c = f \psi$. Then, the Lagrangian takes the form of 
\begin{equation}
    \mathcal{L} = \mathcal{L}[f, A - b] + \mathcal{L}[\psi, b],
\end{equation}
where $A$ is the background field and $b$ is a $\mathrm{U}(1)$ gauge field. We adjust the flux of $b$ so that $f$ sees zero flux, putting $\psi$ into the Laughlin state at filling $\nu_\psi = - 1 / M$. This gives 
\begin{equation}
    \mathcal{L} = \mathcal{L}[f, A - b] + \frac{M}{4 \pi} \beta d \beta + \frac{1}{2 \pi} \beta d b,
\end{equation}
where $\beta$ is a $\mathrm{U}(1)$ gauge field. Next, we put $f$ into a $p + ip$ superconductor \cite{shi2025dopinglattice}, which is described by \cite{ma2020emergent}
\begin{equation}
    \mathcal{L}[f, A - b] = - \frac{2}{4 \pi} \Tr(c d c + \frac{2}{3} c^3) + \frac{1}{4 \pi} \Tr(c) d \Tr(c) + \frac{1}{2 \pi} \Tr(c) d (A - b) - \mathrm{CS}[A - b, g],
\end{equation}
where $c$ is a $\mathrm{U}(2)$ gauge field and $\mathrm{CS}[A - b, g]$ is the Chern-Simons term fo the spin$_\mathbb{C}$ connection. The total Lagrangian takes the form 
\begin{equation}
    \mathcal{L} = - \frac{2}{4 \pi} \Tr(c d c + \frac{2}{3} c^3) + \frac{1}{4 \pi} \Tr(c) d \Tr(c) + \frac{1}{2 \pi} \Tr(c) d (A - b) - \mathrm{CS}[A - b, g] + \frac{M}{4 \pi} \beta d \beta + \frac{1}{2 \pi} \beta d b.
\end{equation}
By integrating out $b$, we obtain 
\begin{equation}
    \mathcal{L} = - \frac{2}{4 \pi} \Tr(c d c + \frac{2}{3} c^3) + \frac{2}{4 \pi} \Tr(c) d \Tr(c) + \frac{M + 1}{4 \pi} \beta d \beta + \frac{1}{2 \pi} \beta d (A - \Tr(c)).
\end{equation}

\subsection{Example: Chiral anyon superconductor from integer quantum Hall state}

The wavefunction of an anyon superconductor obtained by doping the charge-$e$ anyons to the integer quantum Hall state at $\nu = 1$ is 
\begin{equation}
    \Psi = (z - z) \int_u (z - u)^{-1} (u - u) \operatorname{Pf}\!\left(\frac{1}{u - u}\right).
\end{equation}
Following the dictionary, we write the effective Chern-Simons theory as 
\begin{equation}
    \mathcal{L} = - \frac{1}{4 \pi} a d a + \frac{1}{2 \pi} A d a - \frac{2}{4 \pi} \Tr(c d c + \frac{2}{3} c^3) + \frac{2}{4 \pi} \Tr(c) d \Tr(c) + \frac{1}{2 \pi} \beta d (a - \Tr(c)).
\end{equation}
By integrating out $\beta$, we get 
\begin{equation}
    \mathcal{L} = - \frac{2}{4 \pi} \Tr(c d c + \frac{2}{3} c^3) + \frac{1}{4 \pi} \Tr(c) d \Tr(c) + \frac{1}{2 \pi} A d \Tr(c),
\end{equation}
which describes the charge-$2e$ superconductor with $c_- = 3 / 2$.

For the wavefunction 
\begin{equation}
    \Psi = (z - z) \int_u (z - u)^{-1} (u - u) \operatorname{Pf}\!\left(\frac{1}{\bar{u} - \bar{u}}\right),
\end{equation}
the effective Chern-Simons theory is 
\begin{equation}
    \mathcal{L} = - \frac{1}{4 \pi} a d a + \frac{1}{2 \pi} A d a + \frac{2}{4 \pi} \Tr(c d c + \frac{2}{3} c^3) - \frac{2}{4 \pi} \Tr(c) d \Tr(c) - \frac{2}{4 \pi} \beta d \beta - \frac{1}{2 \pi} \beta d (a - \Tr(c)).
\end{equation}
By integrating out $a$, we get 
\begin{equation}
    \mathcal{L} = \frac{2}{4 \pi} \Tr(c d c + \frac{2}{3} c^3) - \frac{2}{4 \pi} \Tr(c) d \Tr(c) - \frac{1}{4 \pi} \beta d \beta + \frac{1}{2 \pi} \beta d (\Tr(c) - A) + \mathrm{CS}[A, g].
\end{equation}
Lastly, by integrating out $\beta$, we get 
\begin{align}
    \mathcal{L} &= \frac{2}{4 \pi} \Tr(c d c + \frac{2}{3} c^3) - \frac{1}{4 \pi} \Tr(c) d \Tr(c) - \frac{1}{2 \pi} \Tr(c) d A + 2 \mathrm{CS}[A, g].
\end{align}
This describes a charge-$2e$ superconductor with $c = 1 / 2$.

\subsection{Example: Chiral anyon superconductor from Laughlin state}

As we discussed in the main text, the wavefunction of an anyon superconductor obtained by doping the charge-$e/3$ anyons to the Laughlin $\nu = 1/3$ state is
\begin{equation}
    \Psi = (z - z)^3 \int_u (z - u)^{-1} \abs{u - u}^{\frac{2}{3}} \int_w (\bar{u} - \bar{w})^{-1} (\bar{w} - \bar{w})^3 \operatorname{Pf}\!\left(\frac{1}{\bar{w} - \bar{w}}\right).
\end{equation}
Following the dictionary, we can write down the effective Chern-Simons theory as 
\begin{equation}
    \mathcal{L} = - \frac{3}{4 \pi} a d a + \frac{1}{2 \pi} A d a + \frac{1}{2 \pi} \hat{b} d a + \frac{2}{4 \pi} \Tr(c d c + \frac{2}{3} c^3) - \frac{2}{4 \pi} \Tr(c) d \Tr(c) + \frac{2}{4 \pi} \beta d \beta - \frac{1}{2 \pi} \beta d (\hat{b} - \Tr(c)),
\end{equation}
where $a$ and $\hat{b}$ are $\mathrm{U}(1)$ gauge fields and $c$ is a $\mathrm{U}(2)$ gauge field. By integrating out $\hat{b}$, we get 
\begin{equation}
    \mathcal{L} = \frac{2}{4 \pi} \Tr(c d c + \frac{2}{3} c^3) - \frac{2}{4 \pi} \Tr(c) d \Tr(c) - \frac{1}{4 \pi} \beta d \beta + \frac{1}{2 \pi} \beta d (A + \Tr(c)),
\end{equation}
Lastly, by integrating out $\beta$, we get 
\begin{align}
    \mathcal{L} &= \frac{2}{4 \pi} \Tr(c d c + \frac{2}{3} c^3) - \frac{1}{4 \pi} \Tr(c) d \Tr(c) + \frac{1}{2 \pi} A d \Tr(c) + \mathrm{CS}[A, g].
\end{align}
This describes a charge-$2e$ superconductor with $c = - 1 / 2$.

For the wavefunction 
\begin{equation}
    \Psi = (z - z)^3 \int_u (z - u)^{-1} \abs{u - u}^{\frac{2}{3}} \int_w (\bar{u} - \bar{w})^{-1} (\bar{w} - \bar{w})^3 \operatorname{Pf}\!\left(\frac{1}{w - w}\right),
\end{equation}
the effective Chern-Simons theory is 
\begin{equation}
    \mathcal{L} = - \frac{3}{4 \pi} a d a + \frac{1}{2 \pi} A d a + \frac{1}{2 \pi} \hat{b} d a - \frac{2}{4 \pi} \Tr(c d c + \frac{2}{3} c^3) + \frac{2}{4 \pi} \Tr(c) d \Tr(c) + \frac{4}{4 \pi} \beta d \beta + \frac{1}{2 \pi} \beta d (\hat{b} - \Tr(c)).
\end{equation}
By integrating out $\hat{b}$, we get 
\begin{equation}
    \mathcal{L} = - \frac{2}{4 \pi} \Tr(c d c + \frac{2}{3} c^3) + \frac{2}{4 \pi} \Tr(c) d \Tr(c) + \frac{1}{4 \pi} \beta d \beta - \frac{1}{2 \pi} \beta d (A + \Tr(c)),
\end{equation}
Lastly, by integrating out $\beta$, we get 
\begin{align}
    \mathcal{L} &= - \frac{2}{4 \pi} \Tr(c d c + \frac{2}{3} c^3) + \frac{1}{4 \pi} \Tr(c) d \Tr(c) - \frac{1}{2 \pi} A d \Tr(c) - \mathrm{CS}[A, g].
\end{align}
This describes a charge-$2e$ superconductor with $c = 1 / 2$.

\subsection{Example: Chiral anyon superconductor from Pfaffian state}

The wavefunction of an anyon superconductor obtained by doping semions to the Pfaffian state is 
\begin{equation}
    \Psi = \operatorname{Pf}\!\left(\frac{1}{z - z}\right) (z - z)^2 \int_u (z - u)^{-1} \abs{u - u} \int_w (\bar{u} - \bar{w})^{-1} (\bar{w} - \bar{w})^2.
\end{equation}
Following the dictionary, we can write down the effective Chern-Simons theory as 
\begin{equation}
    \mathcal{L} = - \frac{2}{4 \pi} \Tr(c d c + \frac{2}{3} c^3) + \frac{2}{4 \pi} \Tr(c) d \Tr(c) - \frac{1}{4 \pi} \beta d \beta + \frac{1}{2 \pi} \beta d (A - \Tr(c)) + \frac{1}{2 \pi} \hat{b} d \beta + \frac{2}{4 \pi} \alpha d \alpha + \frac{1}{2 \pi} \hat{b} d \alpha.
\end{equation}
By integrating out $\beta$, we get 
\begin{equation}
    \mathcal{L} = - \frac{2}{4 \pi} \Tr(c d c + \frac{2}{3} c^3) + \frac{3}{4 \pi} \Tr(c) d \Tr(c) - \frac{1}{2 \pi} A d \Tr(c) + \frac{2}{4 \pi} \alpha d \alpha + \frac{1}{4 \pi} \hat{b} d \hat{b} - \frac{1}{2 \pi} \hat{b} d (\Tr(c) - A - \alpha) + \mathrm{CS}[A, g].
\end{equation}
By integrating out $\hat{b}$, we get 
\begin{equation}
    \mathcal{L} = - \frac{2}{4 \pi} \Tr(c d c + \frac{2}{3} c^3) + \frac{2}{4 \pi} \Tr(c) d \Tr(c) + \frac{1}{4 \pi} \alpha d \alpha - \frac{1}{2 \pi} \alpha d (A - \Tr(c)).
\end{equation}
Lastly, integrating out $\alpha$, we get 
\begin{equation}
    \mathcal{L} = - \frac{2}{4 \pi} \Tr(c d c + \frac{2}{3} c^3) + \frac{1}{4 \pi} \Tr(c) d \Tr(c) + \frac{1}{2 \pi} A d \Tr(c) - \mathrm{CS}[A, g]/
\end{equation}
This describes a charge-$2e$ superconductor with $c = 1 / 2$.

\section{\texorpdfstring{$K$}{K} matrices of Abelian anyon superconductors}

Anyon superconductors emerged from doped Abelian quantum Hall states can be described by the $K$-matrix formulation \cite{wen1992classification}. In the following, we will present the $K$ matrices and charge vectors of the Abelian anyon superconductors discussed in the main text and basis transformations that reveal their superconducting properties explicitly.

\subsection{Semion state}

The anyon superconductor from doped semion state can be described by 
\begin{equation}
    K = 
    \begin{pmatrix}
        2 & -1 & 0 \\
        -1 & 0 & 1 \\
        0 & 1 & -2
    \end{pmatrix}, \quad 
    \vec{t} = 
    \begin{pmatrix}
        1 \\ 0 \\ 0
    \end{pmatrix}.
\end{equation}
The $K$ matrix has a zero eigenvalue and is singular, signaling the existence of a superfluid mode. To see this explicitly, we introduce 
\begin{equation}
    W = 
    \begin{pmatrix}
        0 & 1 & 1 \\
        1 & 0 & 2 \\
        1 & 0 & 1
    \end{pmatrix},
\end{equation}
so that 
\begin{equation}
    W^\mathsf{T} K W = 
    \begin{pmatrix}
        0 & -1 & 0 \\
        -1 & 2 & 0 \\
        0 & 0 & 0 
    \end{pmatrix}, \quad 
    W^\mathsf{T} \vec{t} = 
    \begin{pmatrix}
        0 \\ 1 \\ 1
    \end{pmatrix}.
\end{equation}
Since the upper $2 \times 2$ block has a unit determinant and zero chiral central charge, it does not contribute to a topological term. After integrating out auxiliary gauge fields, the resulting Lagrangian takes the form of 
\begin{equation}
    \mathcal{L} = \frac{1}{2 \pi} A d \alpha + \cdots,
\end{equation}
where the dots represents nontopological terms. This describes a nonchiral charge-$1$ superfluid.

\subsection{\texorpdfstring{$2/3$}{2/3} state}

The $K$ matrix and charge vector of the anyon superconductor from the doped $2 / 3$ state are 
\begin{equation}
    K = 
    \begin{pmatrix}
        1 & 1 & 0 & 0 & -1 \\
        1 & -2 & 0 & 0 & 0 \\
        0 & 0 & -2 & 1 & 1 \\
        0 & 0 & 1 & -2 & 0 \\
        -1 & 0 & 1 & 0 & 0
    \end{pmatrix}, \quad 
    \vec{t} = 
    \begin{pmatrix}
        1 \\ 0 \\ 0 \\ 0 \\ 0
    \end{pmatrix}.
\end{equation}
Introduce the $\mathrm{SL}(5, \mathbb{Z})$ matrix 
\begin{equation}
    W = 
    \begin{pmatrix}
        1 & 0 & 0 & 0 & 2 \\
        0 & 1 & 0 & 0 & 1 \\
        0 & 0 & 1 & 0 & 2 \\
        0 & 0 & 0 & 1 & 1 \\
        0 & 0 & 0 & 2 & 3
    \end{pmatrix},
\end{equation}
such that 
\begin{equation}
    W^\mathsf{T} K W = 
    \begin{pmatrix}
        1 & 1 & 0 & -2 & 0 \\
        1 & -2 & 0 & 0 & 0 \\
        0 & 0 & -2 & 3 & 0 \\
        -2 & 0 & 3 & -2 & 0 \\
        0 & 0 & 0 & 0 & 0 
    \end{pmatrix}, \quad 
    W^\mathsf{T} \vec{t} = 
    \begin{pmatrix}
        1 \\ 0 \\ 0 \\ 0 \\ 2
    \end{pmatrix}.
\end{equation}
The upper $4 \times 4$ block represents a trivial anyon theory with chiral central charge $c = -2$. Therefore, the resulting state is a charge-$2$ superconductor with $c = -2$.

\section{Neutralizing background charge and Gaussian factor}

In this section, we discuss how to determine the Gaussian factors for the parent state wavefunctions and the pseudo wavefunctions. In our formalism, each of the wavefunctions is expressed as the CFT correlator of a string of vertex operators. When we compute the CFT correlators, we used the key identity \cite{francesco1996conformal}
\begin{equation}
    \left\langle \normord{e^{i \alpha_1 \phi(z_1)}} \cdots \normord{e^{i\alpha_n \phi(z_n)}} \right\rangle = \prod_{i < j}^n (z_i - z_j)^{\alpha_i \alpha_j}
\end{equation}
provided the neutrality condition is satisfied:
\begin{equation} \label{eq:neutrality}
    \alpha_1 + \cdots + \alpha_n = 0.
\end{equation}
Here, $\phi$ is a chiral boson which satisfies
\begin{equation}\label{eq:diag_correlator}
    \left\langle \phi(z) \phi(z')\right\rangle = - \log(z - z').
\end{equation}
Eq.~\eqref{eq:neutrality} follows from the $\mathrm{U(1)}$ symmetry $\phi(z) \to \phi(z) + a$ for a constant $a$. We note that, in the current and the next section, we shall use the diagonal basis satisfying Eq.~\eqref{eq:diag_correlator} rather than the $K$ matrix basis introduced in the previous section. We note that the following arguments still hold after we perform the basis transformation Eq.~\eqref{eq:Kmatrix_basis_transf}.  

Therefore, for the CFT correlators in the main text to be non-vanishing, we introduce a uniform neutralizing background charge distribution, following \cite{moore1991nonabelions}. For simplicity, we assume that correlators contain chiral bosons only, as the generalization to correlators containing antichiral bosons would be straightforward. A typical CFT correlator is given by
\begin{equation} \label{eq:general_correlator}
    \left \langle 
    \prod_{\alpha = 1}^{N_0} e^{i q_{a_0}\phi_a \left(u_\alpha^{(0)}\right)} 
    \cdots 
    \prod_{\alpha = 1}^{N_i} e^{i q_{a_i}\phi_a \left(u_\alpha^{(i)}\right)}
    \cdots
    \prod_{\alpha = 1}^{N_m} e^{i q_{a_m}\phi_a \left(u_\alpha^{(m)}\right)}
    \mathcal{O}_{\text{bg},a}
    \right \rangle,
\end{equation}
where $a = 0, 1, \ldots, n$ label the chiral boson species, and $i = 0, 1, \ldots, m$ label the quasiparticle species with $u^{(0)} = z$ denoting the electron coordinates (we henceforth omit the normal ordering symbols.) Here, the neutralizing background charge $\mathcal{O}_{\text{bg},a}$ is given by
\begin{equation}
    \mathcal{O}_{\text{bg},a} = \exp \left[-i \rho_a \int_D d^2 z' \, \phi_a (z')\right],
\end{equation}
where $D$ denotes the region of the circular droplet. Then the correlator Eq.~\eqref{eq:general_correlator} is invariant under $\mathrm{U(1)}$ symmetry $\phi(z) \to \phi(z) + a$ only if
\begin{equation}
    \rho_a \int d^2 z' = \rho_a \pi R^2 = \sum_{i = 1}^{m} q_{a_i} N_i  
    \quad \Rightarrow \quad
    \rho_a = \frac{\sum_{i = 1}^{m} q_{a_i} N_i}{\pi R^2}.
\end{equation}
where $R$ denotes the radius of the droplet. This equation determines each of the uniform background charge density $\rho_a$. 

For a vertex operator $e^{i q_{a_i} \phi_a \left(u_\alpha^{(i)}\right)}$, the contraction with $\mathcal{O}_{\text{bg},a}$ gives
\begin{equation}
    \exp \left[-q_{a_i} \rho_a \int_D d^2z' \, \log(z - z')\right].
\end{equation}
Using the formula
\begin{equation}
    \int_{D} d^2z' \, \ln |z - z'| = \frac{\pi}{2}|z|^2
\end{equation}
up to a boundary term, the real part gives the Gaussian factor
\begin{equation}
    \exp \left[-\frac{\pi}{2} q_{a_i} \rho_a |u_\alpha^{(i)}|^2 \right] 
    = \exp \left[-\frac{q_{a_i} \sum_{j = 1}^{m} q_{a_j} N_j}{2R^2} |u_\alpha^{(i)}|^2 \right].
\end{equation}
The imaginary part can be removed by an appropriate singular gauge transformation as Moore and Read \cite{moore1991nonabelions} suggested. The singular gauge term can also be dealt with flux-tube regularization suggested in \cite{hansson2007composite-fermion}. Thus, the Gaussian factor for $i$-th quasiparticle species arising from the $a$-th chiral boson is given by
\begin{equation}
    \prod_{\alpha = 1}^{N_i} \exp \left[ -\frac{q_{a_i} (\vec{q}_a^\mathsf{T} \vec{N})}{2R^2} |u_{\alpha}^{(i)}|^2 \right],
\end{equation}
where we have defined
\begin{equation}
    \vec{q_{a}} = (q_{a_0}, q_{a_1}, \ldots, q_{a_m})^\mathsf{T}, \qquad \vec{N} = (N_0, N_1, \ldots, N_m)^\mathsf{T}.
\end{equation}
Considering all the chiral boson species, the total Gaussian factor for $i$-th species in the conformal block is given by
\begin{equation}\label{eq:gaussian_factor}
    \prod_{\alpha = 1}^{N_i} \exp \left[ - \sum_{a = 0}^{n} \frac{q_{a_i} (\vec{q}_a^\mathsf{T} \vec{N})}{2R^2} |u_{\alpha}^{(i)}|^2 \right].
\end{equation}
Indeed, this Gaussian factor is independent of the basis for the chiral (and antichiral) bosons we use. Expanding the conformal block Eq.~\eqref{eq:general_correlator}, one can check that it contributes to $M_K$ by $M_K^{(a)} = 2 \vec{q}_a \vec{q}_a^\mathsf{T}$, where $\vec{q}_a \coloneq (q_{a_0}, \cdots, q_{a_m})^\mathsf{T}$. As a result, the full $M_K$ is given by
\begin{equation}\label{eq:M_K_qq}
    M_K = \sum_{a = 0}^{n} M_K^{(a)} = 2 \sum_{a = 0}^{n} \vec{q}_a \vec{q}_a^\mathsf{T}.
\end{equation}
Therefore, the plasma charge vector, whose component is the coefficient of $|u_\alpha^{(i)}|^2$ in the exponential multiplied by a factor or 2, can be expressed as
\begin{equation} \label{eq:plasma_charge_vector}
    \vec{t}_p = -\frac{1}{R^2} \sum_{a = 0}^{n} \vec{q}_a (\vec{q}_a^\mathsf{T} \vec{N}) = -\frac{1}{2R^2} M_K \vec{N} \propto M_K \vec{N}.
\end{equation}
Since $M_K$ is already given in the main text, we can list the corresponding plasma charge vectors $\vec{t}_p$ in Table~\ref{tab:plasma_charge_vectors}, omitting the overall factors of $-1/R^2$. Moreover, $N_i$ denotes the number of particles with coordinates $u^{(i)}$. For wavefunctions containing a Pfaffian factor, however, the contribution of the Pfaffian sector to $M_K$ should be removed before applying Eq.~\eqref{eq:plasma_charge_vector}. This is because $\psi$ is electrically neutral and therefore does not contribute to the plasma charge vector.
\begin{table*}[t]
\centering
\caption{Plasma charge vectors for the different hierarchy constructions, as determined from the Gaussian factors. The entries in each column specify the combinations of particle numbers $N_i$, where $N_0$ denotes the number of physical electrons and $N_i$ the number of $u^{(i)}$ particles.}
    \begin{tabular*}{\textwidth}{@{\extracolsep{\fill}}ccccc@{}}
        \hline
        Laughlin's semion case & $\nu = 2/3$ state & \makecell{Doping $e/3$-anyons \\to Laughlin state} & \makecell{Doping semions \\to Pfaffian} & \makecell{Electron ``condensate'' \\ in $\nu = 1$ state} \\
        \hline 
        $\begin{pmatrix}
            2N_1 -N_2 \\
            -N_1 + N_2 - N_3 \\
            -N_2 + 2N_3
        \end{pmatrix}$ & 
        $\begin{pmatrix}
            N_1 + N_2 - N_3 \\
            N_1 + 4N_2 - 2N_3 \\
            -N_1 - 2 N_2 + 2 N_3 - N_4 \\
            -N_3 + 2N_4 - N_5 \\
            -N_4 + 2N_5
        \end{pmatrix}$ &
        $\begin{pmatrix}
            3N_1 - N_2 \\
            -N_1 + \frac{2}{3} N_2 - N_3 \\
            -N_2 + 3N_3
        \end{pmatrix}$ &
        $\begin{pmatrix}
            2N_1 - N_2 \\
            -N_1 + N_2 - N_3 \\
            -N_2 + 2N_3
        \end{pmatrix}$ &
        $\begin{pmatrix}
                N_1 - N_2 \\
                -N_1 + N_2
            \end{pmatrix}$ 
        \\
        \hline
\end{tabular*}\label{tab:plasma_charge_vectors}
\end{table*}

\section{Orthogonality between \texorpdfstring{$\vec{v}_0$ and $\vec{t}_p$}{}}
In this section, we prove that any null vector $\vec{v}_0$ of the exponent matrix $M_K$ associated with a hierarchy wavefunction is necessarily orthogonal to the plasma charge vector $\vec{t}_p$. As described in the main text, our hierarchy wavefunction is constructed by expressing the parent-state wavefunction and the pseudo-wavefunctions as conformal blocks, multiplying these blocks, and integrating over the anyon coordinates \cite{bonderson2008fractional}. Within this construction scheme, we can prove that $\vec{v}_0^\mathsf{T} \vec{t}_p = 0$. If $\vec{v}_0$ is orthogonal to $\vec{t}_p$, then the density fluctuation along the direction associated to $\vec{v}_0$ does not cost the effective plasma energy $\mathcal{H}[\vec{\rho}]$, implying the existence of a gapless compressible mode \cite{degail2008plasma}.

According to Eq.~\eqref{eq:M_K_qq}, $M_K$ is always positive-semidefinite. Therefore, if $\vec{v}_0$ is a null vector of $M_K$, then
\begin{equation}
    \vec{q}_a^\mathsf{T} \vec{v}_0 = 0 \qquad \forall a = 0, 1, \ldots, n.
\end{equation}

Next, the plasma charge vector can be expressed as Eq.~\eqref{eq:plasma_charge_vector}. Because $\vec{q}_a^\mathsf{T} \vec{v}_0 = 0$, substituting Eq.~\eqref{eq:M_K_qq} into Eq.~\eqref{eq:plasma_charge_vector}, we have
\begin{equation}
    \vec{t}_p^\mathsf{T} \vec{v}_0 \propto \sum_{a = 0}^{n} (\vec{q}_a^\mathsf{T} \vec{v}_0) (\vec{q}_a^\mathsf{T} \vec{N}) = 0.
\end{equation}


\section{Connection to Laughlin's anyon superconductor}
\subsection{Anyon superfluid from semion state}
In this section, we show that the pseudo wavefunction in the case of anyon superfluid from semion state, namely
\begin{equation}\label{eq:semion_pseudo_wf}
    \Phi_1^* = \prod_{\alpha<\beta} \left(\frac{u_\alpha - u_\beta}{|u_\alpha - u_\beta|}\right)^{1/2} \prod_{\alpha < \beta} \frac{\bar{u}_\alpha - \bar{u}_\beta}{|u_\alpha - u_\beta|^{1/2}} 
    \int_{w}\prod_{\alpha, \beta} (\bar{u}_\alpha - \bar{w}_\beta)^{-1} \prod_{\alpha < \beta} (\bar{w}_\alpha - \bar{w}_\beta)^2,
\end{equation}
is equivalent to the state of composite fermions filling two lowest Landau levels. Taking complex conjugate on Eq.~\eqref{eq:semion_pseudo_wf}, we have
\begin{equation}
    \Phi_1 = \prod_{\alpha<\beta} \left(\frac{\bar{u}_\alpha - \bar{u}_\beta}{|u_\alpha - u_\beta|}\right)^{1/2} \prod_{\alpha < \beta} \frac{u_\alpha - u_\beta}{|u_\alpha - u_\beta|^{1/2}} 
    \int_{w}\prod_{\alpha, \beta} (u_\alpha - w_\beta)^{-1} \prod_{\alpha < \beta} (w_\alpha - w_\beta)^2.
\end{equation}
The first factor corresponds to the statistical phase factor, and the remaining factor is a fermionic wavefunction \cite{laughlin1988superconducting}. The fermionic part is characterized by the following $K$ matrix and the charge vector:
\begin{equation}\label{eq:semion_K_matrix}
    K = \begin{pmatrix}
        1 & -1 \\
        -1 & 2
    \end{pmatrix},
    \quad
    \vec{t} = 
    \begin{pmatrix}
        1 & 0
    \end{pmatrix}^\mathsf{T}.
\end{equation}
Now, we shall show that up to the local correlation factor $\prod |u - u|^{-1/2}$ the fermionic wavefunction is equivalent to the state of composite fermions filling two lowest Landau levels:
\begin{equation}
    \Phi \equiv \prod_{\alpha<\beta}(u_\alpha - u_\beta)
    \int_{w}\prod_{\alpha, \beta} (u_\alpha - w_\beta)^{-1} \prod_{\alpha < \beta} (w_\alpha - w_\beta)^2
    \sim
    \psi_{\text{CF}}^{\text{2LL}}(\{u_i\}_{i = 1}^{N_1})
\end{equation}
To show this relation, we express $\Phi$ in terms of a CFT correlator \cite{bonderson2008fractional}:
\begin{equation}\label{eq:pseudo_wavefunction_correlator}
    \Phi = \int_{w} \left \langle \prod_{\alpha} V_1(u_{\alpha}) \prod_{\beta} H_1^{-1} (\omega_\beta) \mathcal{O}_{\text{bg}}\right \rangle,
\end{equation}
where
\begin{equation}
\begin{split}
    V_1(u) &= e^{i \phi_1(u) - i \phi_2 (u)}, \\
    H_1^{-1}(w) &= e^{-i \phi_1(w)}.
\end{split}
\end{equation}
The chiral bosons $\phi_I$ satisfy
\begin{equation}
     \langle \phi_I (u) \phi_J (w) \rangle = - (K^{-1})_{IJ} \log(u - w),
\end{equation}
where $K$ is given in Eq.~\eqref{eq:semion_K_matrix}. Following \cite{suorsa2011general, suorsa2011quasihole}, we replace the quasielectron operator $H^{-1}_1$ with quasilocal quasielectron operator
\begin{equation}
    \mathcal{P}(\bar{w}) = \int d^2 \eta e^{-(1/4) (|\eta|^2 + |w|^2 - 2 \bar{w} \eta)} (H^{-1} \bar{\partial} J)_{\text{gn}}(\eta),
\end{equation}
where `gn' denotes the generalized normal ordering defined in \cite{hansson2009quantum}. Then the pseudo wavefunction is given by \cite{suorsa2011quasihole}
\begin{equation}
    \int [d^2 w_i] \left \langle \prod_{i = 1}^{M} \widetilde{\mathcal{P}}(\bar{w}_i) \prod_{j = 1}^{2M}V_1(u_j) \right\rangle 
    \propto 
    \mathcal{A}
    \left\langle 
    \prod_{i = 1}^M V_1(u_j) 
    \prod_{j = M + 1}^{2M} V_2 (u_j)
    \right\rangle,
\end{equation}
where $M = N/2$, and $N$ is the number of $u$-particles. Here,
\begin{equation}
    V_2(u) = \partial_u \normord{e^{i \phi_2(u)}}
\end{equation}
and $\widetilde{\mathcal{P}}$ is the operator $\mathcal{P}$ adjoined with $e^{2i \phi_2}$, which gives an appropriate pseudo wavefunction.

We have shown that the pseudo wave function $\Phi$ can be equivalently written as,
\begin{equation}
    \Phi \sim     
    \mathcal{A}
    \left\langle 
    \prod_{i = 1}^M V_1(z_j) 
    \prod_{j = M + 1}^{2M} V_2 (z_j)
    \mathcal{O}_{\text{bg}}
    \right\rangle,
\end{equation}
where $V_1(z) = \normord{e^{i (\varphi_1(z) - \varphi_2(z))}}$ and $V_2(z) = \partial_z \normord{e^{i \varphi_2(z)}}$ in the $K$-matrix basis. Previously, we omit the background charge $\mathcal{O}_{\text{bg}}$, but $\mathcal{O}_{\text{bg}}$ plays an important role here. We now claim that this is in fact the determinant of a matrix (up to a multiplicative constant):
\begin{equation}
\begin{split}
    &\mathcal{A}
    \left\langle 
    \prod_{i = 1}^M V_1(z_j) 
    \prod_{j = M + 1}^{2M} V_2 (z_j)
    \mathcal{O}_{\text{bg}}
    \right\rangle \\
    &\propto e^{-\frac{1}{4 l^2}\sum_{i = 1}^{2M} |z_i|^2}
    \begin{vmatrix}
        1 & z_1 & \cdots & z_1^{M-1} & -\frac{1}{4 l^2}\bar{z}_{1} & 1 - \frac{1}{4l^2}\bar{z}_1 z_1 & \cdots & (M-1)z_1^{M-2} - \frac{1}{4l^2} \bar{z}_1 z_1^{M-1} \\
        1 & z_2 & \cdots & z_2^{M-1} & -\frac{1}{4 l^2}\bar{z}_2 & 1 - \frac{1}{4 l^2}\bar{z}_2 z_2 & \cdots & (M-1)z_2^{M-2} - \frac{1}{4 l^2} \bar{z}_2 z_2^{M-1} \\
        \vdots & \vdots & & \vdots & \vdots & \vdots & & \vdots \\
        1 & z_{2M} & \cdots & z_{2M}^{M-1} & -\frac{1}{4 l^2}\bar{z}_{2M} & 1 - \frac{1}{4 l^2}\bar{z}_{2M} z_{2M} & \cdots & (M-1)z_{2M}^{M-2} - \frac{1}{4 l^2} \bar{z}_{2M} z_{2M}^{M-1} 
    \end{vmatrix} \\
    &\propto e^{-\frac{1}{4 l^2}\sum_{i = 1}^{2M} |z_i|^2}
    \begin{vmatrix}
        1 & z_1 & \cdots & z_1^{M-1} & \bar{z}_1 & \bar{z}_1 z_1 & \cdots & \bar{z}_1 z_1^{M-1} \\
        1 & z_2 & \cdots & z_2^{M-1} & \bar{z}_2 & \bar{z}_2 z_2 & \cdots & \bar{z}_2 z_2^{M-1} \\
        \vdots & \vdots & & \vdots &\vdots & \vdots & & \vdots \\
        1 & z_{2M} & \cdots & z_{2M}^{M-1} & \bar{z}_{2M} & \bar{z}_{2M} z_{2M} & \cdots & \bar{z}_{2M} z_{2M}^{M-1}
    \end{vmatrix}
\end{split}
\end{equation}
From the second to the third line, we subtract the first $M$ columns appropriately from the last $M$ columns and then multiply an overall constant to each column. Also, we have set $N = 2M$.

Let $I_1 = \{1, \ldots, M\}$ and $I_2 = \{M+1, \ldots, 2M\}$. Including the contribution from the background charge, the correlator is given by
\begin{equation} \label{eq:Phi_explicit}
    \Phi \sim \mathcal{A} \left[ 
    \prod_{i < j \in I_1} (z_\alpha - z_\beta) \prod_{k \in I_2} \partial_{z_k} \left(\prod_{\alpha < \beta \in I_2} (z_\alpha - z_\beta)  e^{-\frac{1}{4 l^2}\sum_{i = 1}^{2M} |z_i|^2}\right)
    \right].
\end{equation}
We note that for an arbitrary function $f(z)$,
\begin{equation}
    \partial_z (f(z) e^{-\frac{1}{4l^2} |z|^2}) = e^{-\frac{1}{4l^2} |z|^2} \left(\partial_z - \frac{\bar{z}}{4l^2}\right)f(z). 
\end{equation}
Let $\lambda \equiv \frac{1}{4 l^2}$. We then have
\begin{equation*}
\begin{split}
    &\prod_{k \in I_2} \partial_{z_k}\left(\prod_{\alpha < \beta \in I_2} (z_\alpha - z_\beta)  e^{-\frac{1}{4 l^2}\sum_{i = 1}^{2M} |z_i|^2}\right) \\
    &= e^{-\lambda\sum_{i = 1}^{2M} |z_i|^2} \prod_{k \in I_2} \left(\partial_{z_k} - \lambda \bar{z}_k\right) \prod_{\alpha < \beta \in I_2} (z_\alpha - z_\beta) \\
    &= e^{-\lambda\sum_{i = 1}^{2M} |z_i|^2} \prod_{k \in I_2} \left(\partial_{z_k} - \lambda \bar{z}_k\right) 
    \begin{vmatrix}
        1 & z_{M + 1} & z_{M + 1}^2 & \cdots & z_{M + 1}^{M-1} \\
        \vdots & \vdots & \vdots & & \vdots \\
        1 & z_{2M} & z_{2M}^2 & \cdots & z_{2M}^{M-1} \\
    \end{vmatrix} \\
    &= e^{-\lambda\sum_{i = 1}^{2M} |z_i|^2}
    \begin{vmatrix}
        - \lambda \bar{z}_{M + 1} & 1 - \lambda \bar{z}_{M + 1} z_{M + 1} & z_{M + 1} - \lambda \bar{z}_{M + 1} z_{M + 1}^2 & \cdots & (M - 1) z_{M + 1}^{M - 2} - \lambda \bar{z}_{M + 1} z_{M + 1}^{M-1} \\
        - \lambda \bar{z}_{k_{M + 2}} & 1 - \lambda \bar{z}_{M + 2} z_{M + 2} & z_{M + 2} - \lambda \bar{z}_{M + 2} z_{M + 2}^2 & \cdots & (M - 1) z_{M + 2}^{M - 2} - \lambda \bar{z}_{M + 2} z_{M + 2}^{M-1} \\
        \vdots & \vdots & \vdots & & \vdots \\
        - \lambda \bar{z}_{2M} & 1 - \lambda \bar{z}_{2M} z_{2M} & z_{2M} - \lambda \bar{z}_{2M} z_{2M}^2 & \cdots & (M - 1) z_{2M}^{M - 2} - \lambda \bar{z}_{2M} z_{2M}^{M-1} \\
    \end{vmatrix} \\ 
    &\equiv e^{-\lambda\sum_{i = 1}^{2M} |z_i|^2} \det B_{I_2}.
    \end{split} 
\end{equation*}
Here, for a given subset $K \subset \{1, \ldots, 2M\}$, $B_K$ is the $M \times M$ matrix whose entry is
\begin{equation}
    (B_K)_{ir} = r z_{k_i}^{r - 1} - \lambda \bar{z}_{k_i} z_{k_i}^r, \quad r = 0, \ldots , M-1, \quad k_i \in K.
\end{equation}
For a given subset $K^c = \{1, \ldots, 2M\} \setminus K$, let us define
\begin{equation}
    A_{K^c} = \prod_{i < j \in K^c} (z_\alpha - z_\beta)=
    \begin{vmatrix}
        1 & z_{k^c_1} & z_{k^c_1}^2 & \cdots & z_{k^c_1}^{M-1} \\
        1 & z_{k^c_2} & z_{k^c_2}^2 & \cdots & z_{k^c_2}^{M-1} \\
        \vdots & \vdots & \vdots & & \vdots \\
        1 & z_{k^c_M} & z_{k^c_M}^2 & \cdots & z_{k^c_M}^{M-1} \\
    \end{vmatrix},
    \quad
    k^c_i \in K^c.
\end{equation}
We can then write Eq.~\eqref{eq:Phi_explicit} as (ignoring the Gaussian factor)
\begin{equation}
    \Phi \sim \mathcal{A} \left[\det A_{I_2^c} \det B_{I_2}\right]
    = \sum_{\pi \in S_{2M}} \mathrm{sgn}(\pi) \det A_{\{\pi(1), \ldots, \pi(M)\}} \det B_{\{\pi(M + 1), \ldots, \pi(2M)\}}.
\end{equation}
However, any permutation $\pi \in S_{2M}$ can be written as
\begin{equation}
    \pi = (\sigma^1_M \otimes \sigma^2_M) \circ \tau_K.
\end{equation}
Here, $\tau_K$ is a separation of $\{1, \ldots, 2M\}$ into two sets of size $M$ in the increasing order. For example, 
\begin{equation*}
    \tau_{\{1, \ldots, M\}}(\{1, \ldots, 2M\}) = (\{1, \ldots, M\} | \{M+1, \ldots, 2M\}).
\end{equation*}
Also, $\sigma^i_M$ is a permutation on a set of $M$ elements. Since interchanging two rows inside a determinant flips the sign, we have
\begin{equation}
\begin{split}
    \Phi &\sim \mathcal{A} \left[\det A_{I_2^c} \det B_{I_2}\right] \\
    &= \sum_{\substack{K \subset \{1, \ldots, 2M\},|K| = M \\ \sigma^1_M, \sigma^2_M \in S_M}} 
    \mathrm{sgn}(\tau_K) \mathrm{sgn}(\sigma^1_M) \mathrm{sgn}(\sigma^2_M) \\
    &\quad \quad\quad\quad\quad\times 
    \mathrm{sgn}(\sigma^1_M) \det A_{\{\tau_K(1), \ldots, \tau_K(M)\}} \mathrm{sgn}(\sigma^2_M) \det B_{\{\tau_K(M + 1), \ldots, \tau_K(2M)\}} \\
    &= \sum_{\substack{K \subset \{1, \ldots, 2M\},|K| = M \\ \sigma^1_M, \sigma^2_M \in S_M}} 
    \mathrm{sgn}(\tau_K) \det A_{\{\tau_K(1), \ldots, \tau_K(M)\}} \det B_{\{\tau_K(M + 1), \ldots, \tau_K(2M)\}} \\
    &\propto \sum_{K \subset \{1, \ldots, 2M\},|K| = M} \epsilon(K) \det A_{\{\tau_K(1), \ldots, \tau_K(M)\}} \det B_{\{\tau_K(M + 1), \ldots, \tau_K(2M)\}}.
\end{split}
\end{equation}
In the second equality, we use the fact that $\mathrm{sgn}^2 = 1$. But using the generalized Laplace expansion, the last expression is exactly (up to sign)
\begin{equation}
    \begin{vmatrix}
        1 & z_1 & \cdots & z_1^{M-1} & -\frac{1}{4}\bar{z}_{M} & 1 - \frac{1}{4}\bar{z}_1 z_1 & \cdots & (M-1)z_1^{M-2} - \frac{1}{4} \bar{z}_1 z_1^{M-1} \\
        1 & z_2 & \cdots & z_2^{M-1} & -\frac{1}{4}\bar{z}_2 & 1 - \frac{1}{4}\bar{z}_2 z_2 & \cdots & (M-1)z_2^{M-2} - \frac{1}{4} \bar{z}_2 z_2^{M-1} \\
        \vdots & \vdots & & \vdots & \vdots & \vdots & & \vdots \\
        1 & z_{2M} & \cdots & z_{2M}^{M-1} & -\frac{1}{4}\bar{z}_{2M} & 1 - \frac{1}{4}\bar{z}_{2M} z_{2M} & \cdots & (M-1)z_{2M}^{M-2} - \frac{1}{4} \bar{z}_{2M} z_{2M}^{M-1} 
    \end{vmatrix}.
\end{equation}
\subsection{Anyon superconductor from 2/3 state}
In this section, we derive an anyon superconductor from the $\nu = 2 / 3$ state. The $\nu = 2 / 3$ state can be viewed as the particle-hole conjugate of the Laughlin $\nu = 1 / 3$ state, or equivalently as a hole condensate on top of the $\nu = 1$ integer quantum Hall state \cite{girvin1984particle-hole}. To separate this condensation from the subsequent quasiparticle condensations in the hierarchy construction, we decompose the $K$ matrix as
\begin{equation}\label{eq:2/3_K_matrix}
    K = 
    \begin{pmatrix}
        1 & 1 & -1 & 0 & 0 \\
        1 & -2 & 0 & 0 & 0 \\
        -1 & 0 & 0 & 1 & 0 \\
        0 & 0 & 1 & -2 & 1 \\
        0 & 0 & 0 & 1 & -2
    \end{pmatrix}
    =
    \begin{pmatrix}
        1 & 1 & -1 & 0 & 0 \\
        1 & 1 & -1 & 0 & 0 \\
        -1 & -1 & 1 & 0 & 0 \\
        0 & 0 & 0 & 0 & 0\\
        0 & 0 & 0 & 0 & 0
    \end{pmatrix}
    -
    \begin{pmatrix}
        0 & 0 & 0 & 0 & 0 \\
        0 & 3 & -1 & 0 & 0 \\
        0 & -1 & 1/3 & 0 & 0 \\
        0 & 0 & 0 & 0 & 0 \\
        0 & 0 & 0 & 0 & 0 \\
    \end{pmatrix}
    -
    \begin{pmatrix}
        0 & 0 & 0 & 0 & 0 \\
        0 & 0 & 0 & 0 & 0 \\
        0 & 0 & 2/3 & -1 & 0 \\
        0 & 0 & -1 & 2 & -1 \\
        0 & 0 & 0 & -1 & 2
    \end{pmatrix}
\end{equation}
The first two matrices on the RHS describe the parent $\nu = 2 / 3$ state with, while the last matrix represents the state formed by the condensate of the charge-$2e/3$ quasiparticles, as we will show shortly. The charge vector is given by
\begin{equation}
    \vec{t} = 
    \begin{pmatrix}
        1 & 0 & 0 & 0 & 0
    \end{pmatrix}^\mathsf{T}.
\end{equation}

As before, we can express the parent $\nu = 2 / 3$ state with the $2e/3$ quasiparticles as
\begin{equation}
    \Psi_0 = 
    \int_{u}
    \left\langle 
    \prod_{\alpha = 1}^{N_0} e^{i \phi_1 (z_\alpha)}
    \prod_{\beta = 1}^{N_1} e^{i \phi_2 (u_\beta)} 
    \prod_{\gamma = 1}^{N_2} e^{i \phi_3 (w_\gamma)} 
    \prod_{j = 1}^{N_1} e^{i \bar{\phi}'_2(\bar{u}_j)} 
    \prod_{k = 1}^{N_2} e^{i \bar{\phi}'_3(\bar{w}_k)}
    \right\rangle.
\end{equation}
The pseudowavefunction of the $2e/3$ anyons is given by
\begin{equation}
    \Phi_1^* = \int_{\xi, \zeta} \left\langle
    \prod_{\alpha = 1}^{N_2} e^{- i \bar{\phi}_3 (\bar{w}_\alpha)}
    \prod_{\beta = 1}^{N_3} e^{- i \bar{\phi}_4 (\bar{\xi}_\beta)} 
    \prod_{\gamma = 1}^{N_4} e^{- i \bar{\phi}_5 (\bar{\zeta}_\gamma)}
    \right\rangle,
\end{equation}
where $\xi_\beta$ and $\zeta_\gamma$ are auxiliary hierarchy coordinates that are integrated over. The chiral bosons $\phi_I$ and the antichiral bosons $\bar{\phi}'$ and $\bar{\phi}$ satisfy
\begin{align}
    \left\langle \phi_I(z) \phi_J(w) \right\rangle
    &= -\kappa_{IJ} \log(z - w), \nonumber\\
    \left\langle \bar{\phi}'_I(z) \bar{\phi}'_J(w) \right\rangle
    &= -\bar{\kappa}'_{IJ} \log(\bar{z} - \bar{w}), \nonumber \\
    \left\langle \bar{\phi}_I(z) \bar{\phi}_J(w) \right\rangle
    &= -\bar{\kappa}_{IJ} \log(\bar{z} - \bar{w}),
\end{align} 
where 
\begin{equation}
    \kappa = 
    \begin{pmatrix}
        1 & 1 & -1 & 0 & 0 \\
        1 & 1 & -1 & 0 & 0 \\
        -1 & -1 & 1 & 0 & 0 \\
        0 & 0 & 0 & 0 & 0\\
        0 & 0 & 0 & 0 & 0
    \end{pmatrix},
    \quad
    \bar{\kappa}' = 
    \begin{pmatrix}
        0 & 0 & 0 & 0 & 0 \\
        0 & 3 & -1 & 0 & 0 \\
        0 & -1 & 1/3 & 0 & 0 \\
        0 & 0 & 0 & 0 & 0 \\
        0 & 0 & 0 & 0 & 0 \\
    \end{pmatrix},
    \quad
    \bar{\kappa} = 
    \begin{pmatrix}
        0 & 0 & 0 & 0 & 0 \\
        0 & 0 & 0 & 0 & 0 \\
        0 & 0 & 2/3 & -1 & 0 \\
        0 & 0 & -1 & 2 & -1 \\
        0 & 0 & 0 & -1 & 2
    \end{pmatrix}.
\end{equation}
We note that these matrices appeared in Eq.~\eqref{eq:2/3_K_matrix}. Expanding the conformal blocks above, we obtain
\begin{equation}
    \Psi_0 = (z - z) \int_u (z - u) \abs{u - u}^2 (\bar{u} - \bar{u})^2 \abs{u - w}^{-2} (z - w)^{-1} \abs{w - w}^{\frac{2}{3}} (w - w)^{\frac{2}{3}}
\end{equation}
and
\begin{equation}\label{eq:2/3_pseudo_wf}
    \Phi_1^* = (\bar{w} - \bar{w})^{\frac{2}{3}} 
    \int_{\xi, \zeta} (\bar{w} - \bar{\xi})^{-1} 
    (\bar{\xi} - \bar{\xi})^2 (\bar{\xi} - \bar{\zeta})^{-1}
    (\bar{\zeta} - \bar{\zeta})^2,
\end{equation}
where we ahve used the shorthand notation for simplicity, for example, $(z - z) \equiv \prod_{\alpha < \beta} (z_\alpha - z_\beta)$. The hierarchy integral is then given by
\begin{align}
    \Psi_1 = \int_{w} \Phi_1^* \Psi_0 = 
    (z - z) \int_u (z - u) \abs{u - u}^2 (\bar{u} - \bar{u})^2 \int_w \abs{u - w}^{-2} (z - w)^{-1} \abs{w - w}^2& \nonumber \\
    \times \int_\xi (\bar{w} - \bar{\xi})^{-1} (\bar{\xi} - \bar{\xi})^2 \int_\zeta (\bar{\xi} - \bar{\zeta})^{-1} (\bar{\zeta} - \bar{\zeta})^2&.
\end{align}
The exponent matrix is
\begin{equation}
    M_K = 
    \begin{pmatrix}
        2 & 2 & -2 & 0 & 0 \\
        2 & 8 & -4 & 0 & 0 \\
        -2 & -4 & 4 & -2 & 0 \\
        0 & 0 & -2 & 4 & -2 \\
        0 & 0 & 0 & -2 & 4
    \end{pmatrix},
\end{equation}
which is positive-semidefinite and has a null vector $\vec{v}_0 = \begin{pmatrix} 2 & 1 & 3 & 2 & 1 \end{pmatrix}^\mathsf{T}$. Since $\vec{v}_0^\mathsf{T} \vec{t} = 2$, the hierarchy state describes a charge-$2e$ superconductor

After taking complex conjugation, the pseudowavefunction Eq.~\eqref{eq:2/3_pseudo_wf} is given by
\begin{align}
    \Phi_1 &= (w - w)^{2/3} \int_{\xi, \zeta} (w - \xi)^{-1} (\xi - \xi)^2 (\xi - \zeta)^{-1} (\zeta - \zeta)^2 \nonumber \\
    &= \left(\frac{\bar{w} - \bar{w}}{|w - w|}\right)^{1/3} 
    \frac{w - w}{|w - w|^{1/3}} \int_{\xi, \zeta} (w - \xi)^{-1} (\xi - \xi)^2 (\xi - \zeta)^{-1} (\zeta - \zeta)^2.
\end{align}
In the last line, the first factor represents a statistical gauge term, while the remaining term describes the fermionic wavefunction. Up to a local correlation factor $|w - w|^{-1/3}$, the fermionic wavefunction can be expressed as
\begin{equation}
    \Phi \equiv (w - w) \int_{\xi, \zeta} (w - \xi)^{-1} (\xi - \xi)^2 (\xi - \zeta)^{-1} (\zeta - \zeta)^2.
\end{equation}
Now, we shall demonstrate that this pseudowavefunction is equivalent to a composite-fermion state with the three lowest Landau levels filled. A $K$ matrix and a charge vector that describes this fermionic wavefunction are
\begin{equation}
    K = 
    \begin{pmatrix}
        1 & -1 & 0 \\
        -1 & 2 & -1 \\
        0 & -1 & 2
    \end{pmatrix},
    \quad
    \vec{t} = 
    \begin{pmatrix}
        1 \\ 0 \\ 0
    \end{pmatrix}.
\end{equation}
As before, we follow \cite{suorsa2011quasihole} to construct the composite fermion operators. In the $K$ matrix basis
\begin{equation}
    \left\langle \phi_I (z) \phi_J (w) \right\rangle
    = - (K^{-1})_{IJ} \log(z - w),
\end{equation}
the lowest Landau level electron operator is given by
\begin{equation}
    V_1(z) = e^{i \phi_1 (z) - i \phi_2 (z)}.
\end{equation}
The quasihole operator is 
\begin{equation}
    H_1(w) = e^{i \phi_1(w)}
\end{equation}
so that the quasielectron operator is given by
\begin{equation}
    P_1(w) = (H^{-1}_1 V_1)_{\text{gn}}(w) = \partial_w e^{- i \phi_2 (w)}.
\end{equation}
To obtain an electronic operator, we fuse $P_1$ with a local auxiliary field as
\begin{equation}
    V_2(w) = P_1(w) e^{2i \phi_2(w) - i\phi_3(z)} = \partial_w e^{i \phi_2 (w) - i \phi_3(w)}.
\end{equation}
The next quasihole operator is 
\begin{equation}
    H_2(w) = e^{i \phi_2 (w)},
\end{equation}
which yields the next quasielectron operator:
\begin{equation}
    P_2(w) = (H_2^{-1} V_2)_{\text{gn}}(w) = \partial_w^2 e^{- i \phi_3(z)}.
\end{equation}
The final electronic operator is then obtained by
\begin{equation}
    V_3(z) = P_2(z) e^{2 i \phi_3 (z)} = \partial_w^2 e^{i \phi_3 (w)}.
\end{equation}
As a result, the hierarchy wavefunction can be equivalently represented as (with $M = N_2$)
\begin{equation}\label{eq:2/3_composite_fermions}
    \Phi = \mathcal{A} \left\langle \prod_{i = 1}^{M/3} V_1(w_i) \prod_{i = M/3 + 1}^{2M/3} V_2(w_i) \prod_{i = 2M/3 + 1}^{M} V_3(w_i) \mathcal{O}_{\text{bg}}\right\rangle,
\end{equation}
which can be understood as a wavefunction filling three lowest Landau levels. We note that the electron operators $V_i$ indeed characterizes physcial electrons since they are represented as
\begin{equation}
    l_1 = 
    \begin{pmatrix}
        1 \\ -1 \\ 0 
    \end{pmatrix},
    \quad
    l_2 = 
    \begin{pmatrix}
        0 \\ 1 \\ -1
    \end{pmatrix},
    l_3 =
    \begin{pmatrix}
        0 \\ 0 \\ 1
    \end{pmatrix},
\end{equation}
which satisfy
\begin{equation}
    l_i K^{-1} l_j = \delta_{ij}, \qquad \vec{t}^\mathsf{T} K^{-1} l_i = 1 \qquad(i = 1, 2 , 3).
\end{equation}

We finally explicitly show that Eq.~\eqref{eq:2/3_composite_fermions} can be represented as the slater determinant corresponding to the state filling three lowest Landau levels. The Gaussian factor due to the background charge operator $\mathcal{O}_{\text{bg}}$ is given by $\exp (-\frac{1}{4 l^2} \sum_{i = 1}^N |w_i|^2)$. Thus, the correlator is given by
\begin{equation}
\begin{split}
    &\prod_{i < j \in I_1} (w_i - w_j) e^{-\frac{1}{4 l^2} \sum_{i \in I_1}|w_i|^2} \\
    &\times 
    \prod_{i \in I_2} \partial_{w_i} \left[\prod_{i < j \in I_2} (w_i - w_j)  e^{-\frac{1}{4 l^2} \sum_{i \in I_2}|w_i|^2} \right]
    \prod_{i \in I_3} \partial^2_{w_i} \left[\prod_{i < j \in I_3} (w_i - w_j)  e^{-\frac{1}{4 l^2} \sum_{i \in I_3}|w_i|^2} \right],
\end{split}
\end{equation}
where $I_1 = \{1, \ldots, N/3\}$, $I_2 = \{N/3 + 1 , \ldots, 2N/3\}$, and $I_3 = \{2N/3 +1 , \ldots, N\}$. As before, we define $\lambda = \frac{1}{4l^2}$ and apply the derivatives to the Gaussian factors to write
\begin{equation}
\begin{split}
    &e^{-\frac{1}{4 l^2} \sum_{i = 1}^N|w_i|^2} \prod_{i < j \in I_1} (w_i - w_j)  \\
    &\times 
    \prod_{i \in I_2} \left(\partial_{w_i} - \lambda \bar{w}_i \right) \prod_{i < j \in I_2} (w_i - w_j) 
    \prod_{i \in I_3} (\partial_{w_i} - \lambda \bar{w}_i)^2\prod_{i < j \in I_3} (w_i - w_j).
\end{split}
\end{equation}
Using the same trick we used before, we can write this as
\begin{equation}
    e^{-\frac{1}{4 l^2}  \sum_{i = 1}^N|w_i|^2} \det A_{I_1} \det B_{I_2} \det C_{I_3},
\end{equation}
where $A_{I}$, $B_{I}$, and $C_I$ are $M \times M$ matrices (here $M = N/3$) whose entries are given by
\begin{equation}
\begin{split}
    (A_I)_{ir} &= w_i^r, \quad i \in I, \\
    (B_I)_{ir} &= r w_i^{r - 1} - \lambda \bar{w}_i w_i, \quad i \in I, \\
    (C_I)_{ir} &= r(r-1) w^{r - 2} - 2 \lambda r \bar{w}_i w_i^{r - 1} + \lambda^2 \bar{w}^2 w^r, \quad i \in I,
\end{split}
\end{equation}
and $0 \leq r \leq M - 1$. Therefore, we have
\begin{equation}
\begin{split}
    \Phi &= \mathcal{A} \left[\det A_{I_1} \det B_{I_2} \det C_{I_3} \right] \\
    &= \frac{1}{N!} \sum_{\pi \in S_{3M}} \mathrm{sgn}(\pi)\det A_{\{\pi(1), \ldots, \pi(M)\}} \det B_{\{\pi(M + 1), \ldots, \pi(2M)\}} \det C_{\{\pi(2M + 1), \ldots, \pi(3M)\}},
\end{split}
\end{equation}
where we have suppressed the Gaussian factor (Note that anti-symmetrizing the Gaussian factor gives just the same Gaussian factor.) But any permutation $\pi \in S_{3M}$ can be decomposed as
\begin{equation}
    \pi = (\sigma^1_M \times \sigma^2_M \times \sigma^3_M) \circ \tau_{3M},
\end{equation}
where $\tau_{3M}$ is a separation of $\{1, \ldots, 3M\}$ in three sets of size $M$ in the increasing order. For example, 
\begin{equation}
    \tau_{3M} (\{1,\ldots, 3M\}) = (\{1, \ldots, M\} | \{M+1, \ldots, 2M\} | \{2M + 1, \ldots, 3M\}).
\end{equation}
Also, each $\sigma^i_M$ permutes the $i$-th set of size $M$. Since permuting the rows of the determinant only flips the sign without changing the absolute value, we have
\begin{equation}
\begin{split}
    \Phi &= \frac{1}{N!} \sum_{\sigma^1_M, \sigma^2_M, \sigma^3_M} \sum_{\tau_{3M}} \big\{ \mathrm{sgn}(\sigma_M^1) \, \mathrm{sgn}(\sigma_M^2) \, \mathrm{sgn}(\sigma_M^3) \, \mathrm{sgn}(\tau_{3M}) \\
    & \qquad  \times \mathrm{sgn}(\sigma_M^1) \det A_{\{\tau(1), \ldots, \tau(M)\}} 
    \mathrm{sgn}(\sigma_M^2) \, \det B_{\{\tau(M + 1), \ldots, \tau(2M)\}}
    \, \mathrm{sgn}(\sigma_M^3) \det C_{\{\tau(2M + 1), \ldots, \tau(3M)\}} \big\} \\
    &= \frac{1}{N!} \sum_{\sigma^1_M, \sigma^2_M, \sigma^3_M} \sum_{\tau_{3M}} \mathrm{sgn}(\tau_{3M}) \det A_{\{\tau(1), \ldots, \tau(M)\}} 
    \det B_{\{\tau(M + 1), \ldots, \tau(2M)\}}
    \det C_{\{\tau(2M + 1), \ldots, \tau(3M)\}} \big\} \\
    &=  \frac{(M!)^3}{N!} \sum_{\tau_{3M}} \mathrm{sgn}(\tau_{3M}) \det A_{\{\tau(1), \ldots, \tau(M)\}} 
    \det B_{\{\tau(M + 1), \ldots, \tau(2M)\}}
    \det C_{\{\tau(2M + 1), \ldots, \tau(3M)\}} 
\end{split}
\end{equation}
But we have
\begin{equation}
\begin{split}
    &\frac{(M!)^3}{N!} \sum_{\tau_{3M}} \mathrm{sgn}(\tau_{3M}) \det A_{\{\tau(1), \ldots, \tau(M)\}} 
    \det B_{\{\tau(M + 1), \ldots, \tau(2M)\}}
    \det C_{\{\tau(2M + 1), \ldots, \tau(3M)\}} \\
    &=\frac{(M!)^3}{N!} \sum_{\tau_{3M}|_{\{1, \ldots, M\}}} \mathrm{sgn}(\tau_{3M}|_{\{1, \ldots, M\}}) \det A_{\{\tau(1), \ldots, \tau(M)\}} \\
    & \times\sum_{\substack{\tau_{3M}|_{\{M + 1, \ldots, 3M\}}  \\ \tau(1), \ldots, \tau(M) \text{ fixed}}} 
    \mathrm{sgn}(\tau_{3M}|_{\{M + 1, \ldots, 3M\}}) \det B_{\{\tau(M + 1), \ldots, \tau(2M)\}}
    \det C_{\{\tau(2M + 1), \ldots, \tau(3M)\}} \\
    &=\frac{(M!)^3}{N!} \sum_{\tau_{3M}|_{\{1, \ldots, M\}}} \mathrm{sgn}(\tau_{3M}|_{\{1, \ldots, M\}}) \det A_{\{\tau(1), \ldots, \tau(M)\}} \times \det((B|C)_{\{1, \ldots, 3M\} \setminus \{\tau(1), \ldots, \tau(M)\}}) \\
    &=\frac{(M!)^3}{N!} \det(A | B | C).
\end{split}
\end{equation}
Here, $(B | C)$ is a square matrix defined by
\begin{equation}
    (B | C)_{I \cup J} = (B_{I \cup J} | C_{I \cup J}).
\end{equation}
$(A | B | C)$ is defined similarly. Therefore, the pseudowavefunction for the $2/3$ case is also proportional to the determinant of the following matrix:
\begin{equation}
    ([w_i^r] | [rw_i^{r - 1} - \lambda \bar{w}_i w_i ] | [r(r-1)w^{r - 2} - 2 \lambda r \bar{w}_i w_i^{r - 1} + \lambda^2 w^2 w^r])_{0 \leq r \leq M -1 , 1 \leq i \leq 3M}
\end{equation}

\section{Off-diagonal long-range order}
\subsection{Construction of the ODLRO operator}
In this section, we explain how to introduce the ODLRO operator for a general hierarchy wavefunction describing a superfluid or a superconductor. 

As described in the main text, a parent wavefunction and a pseudowavefunction are expressed in terms of vertex operators. In particular, the parent wavefunction contains chiral bosons and the pseudowavefunction contains antichiral bosons. We first combine them into a single CFT correlator. For each particle species (for example, the $z$-, $u$-, and $w$-species in the semion case), we then collect all vertex operators associated with the same particle coordinate and identify their product as the full vertex operator associated with that particle species. We note that the resulting vertex operators may contain both chiral and antichiral fields. 

For example, in the semion case, the $u$ coordinate appears in vertex operators $e^{i \phi_2(u)}$ in the parent wavefunction sector and in vertex operators $e^{-i \bar{\phi}_2(\bar{u})}$ in the pseudowavefunction sector. We therefore identify the $u$-particle operator as $V_u (u, \bar{u}) = e^{i \phi_2(u) - i \bar{\phi}_2(\bar{u})}$. Similarly, the $z$- and $w$-particle operators are given by $V_z(z) = e^{i \phi_1(z)}$ and $V_w (\bar{w}) = e^{-i \bar{\phi}_3 (\bar{w})}$.

Once the particle operator for each species has been identified, an operator creating a composite object labeled by $\vec{v} = (v_1, \ldots, v_m)$ can be constructed as follows. Suppose that there are $m$ species of particles labeled by $u^{(\alpha)}$, $\alpha = 1, \ldots, m$, and the corresponding particle operators are given by
\begin{equation}
    V_{\alpha}(u^{(\alpha)}, \bar{u}^{(\alpha)}) = e^{i \sum_{I} q_{\alpha I} \phi_{I}(u^{(\alpha)}) + i \sum_{I} \bar{q}_{\alpha I} \bar{\phi}_I (\bar{u}^{(\alpha)})}, \quad \alpha = 1, \ldots, m.
\end{equation}
For convenience, we introduce the combinations $\varphi_\alpha(u^{(\alpha)}, \bar{u}^{(\alpha)}) =  \sum_{I} q_{\alpha I} \phi_{I}(u^{(\alpha)}) +  \sum_{I} \bar{q}_{\alpha I} \bar{\phi}_I (\bar{u}^{(\alpha)}), \quad \alpha = 1, \ldots, m$. Then the operator creating the composite object $\vec{v}$ at position $\eta$ is defined by
\begin{equation}
    \mathcal{O}_{\vec{v}}(\eta, \bar{\eta}) \sim \exp{i \vec{v}^\mathsf{T} \vec{\varphi}(\eta, \bar{\eta})} = \exp{i (v_1 \varphi_1 (\eta, \bar{\eta}) + \cdots + v_m \varphi_m (\eta, \bar{\eta}))},
\end{equation}
where $\vec{\varphi} = (\varphi_1, \ldots, \varphi_m)^\mathsf{T}$. We note that this operator should be understood with appropriate short-distance regularization, such as point splitting. When $\vec{v}$ is an integer null vector of the exponent matrix $M_K$, the operator $\mathcal{O}_{\vec{v}}$ defines the corresponding ODLRO operator, denoted by $\mathcal{O}_{\text{SC}}(\eta)$ in the main text.

The above discussion explains how the ODLRO operators are constructed in the semion case and in the state obtained by doping e/3-anyons into the Laughlin state. The same construction can be applied to other cases. For example, for the $\nu = 2/3$ case, the ODLRO operator is given by
\begin{equation}
    \mathcal{O}_{\text{SC}} (\eta) \sim \exp{i [2\phi_1(\eta) + (\phi_2(\eta) + \bar{\phi}'_2(\bar{\eta})) + 3 (\phi_3 (\eta) + \bar{\phi}'_3 (\bar{\eta}) - \bar{\phi}_3 (\bar{\eta})) - 2 \bar{\phi}_4 (\bar{\eta}) - \bar{\phi}_5(\bar{\eta})]}.
\end{equation}

\subsection{Relation between the ODLRO operator and the reduced density matrix}
In this section, we explicitly show how the ODLRO operator defined above is related to the reduced density matrix considered in the main text. We focus on the semion case for concreteness, while the same argument applies to the other cases. In the main text, we argue that ODLRO is diagnosed through the reduced density matrix
\begin{equation}
    \rho(\eta, \eta') = \frac{\int_z \Psi_1^*(\eta; \{z_i\}) \Psi_1(\eta'; \{z_i\})}{\int_z \Psi_1^*(\{z_i\}) \Psi_1(\{z_i\})}. \nonumber
\end{equation}
Here,
\begin{align}\label{eq:odlro_wf}
    &\Psi_{1}(\eta; \{z_i\}) \nonumber \\
    &= \int_{u,w} \left\langle
    \mathcal{O}_{\text{SC}}(\eta)
    \prod_{\alpha = 1}^{N_1}
    e^{i \phi_1(z_\alpha)}
    \prod_{\beta = 1}^{N_2}
    e^{i \phi_2(u_\beta) - i \bar{\phi}_2(\bar{u}_\beta)}
     \prod_{\gamma = 1}^{N_3}
    e^{- i \bar{\phi}_3(\bar{w}_\gamma)}
    \right\rangle.
\end{align} 
For convenience, we combine the chiral and antichiral bosons into a single CFT correlator. This does not alter the resulting hierarchy wavefunction, since the chiral and the antichiral fields commute with one another. The ODLRO operator $\mathcal{O}_{\text{SC}}(\eta)$ is defined as 
\begin{equation}\label{eq:odlro}
    \mathcal{O}_{\text{SC}}(\eta) \sim \exp{i \phi_1(\eta) + 2 i [\phi_2(\eta) - \bar{\phi}_2(\bar{\eta})] -i\bar{\phi}_3(\bar{\eta})}.
\end{equation}
We now show explicitly how inserting $\mathcal{O}_{\text{SC}}(\eta)$ into the conformal block leads to the conventional reduced density matrix.

First, this definition satisfies 
\begin{align}\label{eq:odlro_property}
    &\left\langle
    \mathcal{O}_{\text{SC}}(\eta)
    \prod_{\alpha = 1}^{N_1}
    e^{i \phi_1(z_\alpha)}
    \prod_{\beta = 1}^{N_2}
    e^{i \phi_2(u_\beta) - i \bar{\phi}_2(\bar{u}_\beta)}
    \prod_{\gamma = 1}^{N_3}
    e^{- i \bar{\phi}_3(\bar{w}_\gamma)}
    \right\rangle \nonumber \\
    &\qquad\qquad\qquad\qquad\qquad=\left\langle
    \prod_{\alpha = 1}^{N_1}
    e^{i \phi_1(z_\alpha)}
    \prod_{\beta = 1}^{N_2}
    e^{i \phi_2(u_\beta) - i \bar{\phi}_2(\bar{u}_\beta)}
    \prod_{\gamma = 1}^{N_3}
    e^{- i \bar{\phi}_3(\bar{w}_\gamma)}
    \right\rangle,
\end{align}
apart from regulator-dependent factors. The contraction of $\mathcal{O}_{\text{SC}}(\eta)$ with $\prod_\alpha e^{i\phi_1(z_\alpha)}$ yields
\begin{equation*}
    \left\langle e^{i \phi_1(\eta) + 2 i [\phi_2(\eta) - \bar{\phi}_2(\bar{\eta})] -i\bar{\phi}_3(\bar{\eta})} \prod_{\alpha}e^{i\phi_1(z_\alpha)} \cdots \right\rangle
    = (\eta - z)^{1 \kappa_{11} + 2 \kappa_{12}} = (\eta - z)^0 = 1,
\end{equation*}
while its contraction with $\prod_{\beta} e^{i\phi_2(u_\beta) - i \bar{\phi}_2(\bar{u}_\beta)}$ yields
\begin{equation*}
    \left\langle e^{i \phi_1(\eta) + 2 i [\phi_2(\eta) - \bar{\phi}_2(\bar{\eta})] -i\bar{\phi}_3(\bar{\eta})} \prod_{\beta} e^{i\phi_2(u_\beta) - i \bar{\phi}_2(\bar{u}_\beta)} \cdots \right\rangle
    = (\eta - u)^{1 \kappa_{21} + 2 \kappa_{22}}(\bar{\eta} -\bar{u})^{2\bar{\kappa}_{22} + 1 \bar{\kappa}_{23}}
    = (\eta - u)^0 (\bar{\eta} -\bar{u})^0 = 1.
\end{equation*}
Here, we have used a shorthand definition $(\eta - z) \equiv \prod_{\alpha} (\eta - z_\alpha)$, and similarly for the other coordinates. Finally, its contraction with $\prod_{\gamma} e^{- i \bar{\phi}_3(\bar{w}_\gamma)}$ yields
\begin{equation*}
    \left\langle e^{i \phi_1(\eta) + 2 i [\phi_2(\eta) - \bar{\phi}_2(\bar{\eta})] -i\bar{\phi}_3(\bar{\eta})} \prod_{\gamma} e^{- i \bar{\phi}_3(\bar{w}_\gamma)} \cdots \right\rangle
    = (\bar{\eta} -\bar{w})^{2\bar{\kappa}_{32} + 1 \bar{\kappa}_{33}}
    = (\bar{\eta} -\bar{w})^0 = 1.
\end{equation*}
Thus, inserting $\mathcal{O}_{\text{SC}}$ does not alter the polynomial part of the hierarchy integrand. This property follows from the fact that
\begin{equation}
    \kappa\vec{v}_0 = 0 \qquad \text{and} \qquad \bar{\kappa} \vec{v}_0 = 0.    
\end{equation}

We next consider a general vector $\vec{v} = (v_1, v_2, v_3)$ and define the corresponding operator
\begin{equation}
    \mathcal{O}_{\vec{v}}(\eta) \sim \exp\left({i v_1 \phi_1(\eta) +  i v_2 [\phi_2(\eta) - \bar{\phi}_2(\bar{\eta})] -i v_3 \bar{\phi}_3(\bar{\eta}) }\right).
\end{equation}
Following the same steps as above, inserting $\mathcal{O}_{\vec{v}}(\eta)$ into the conformal block yields
\begin{align}
    &\left\langle e^{i v_1 \phi_1(\eta) + i v_2 [\phi_2(\eta) - \bar{\phi}_2(\bar{\eta})] -i v_3 \bar{\phi}_3(\bar{\eta})} \prod_{\alpha = 1}^{N_1}
    e^{i \phi_1(z_\alpha)}
    \prod_{\beta = 1}^{N_2}
    e^{i \phi_2(u_\beta) - i \bar{\phi}_2(\bar{u}_\beta)}
    \prod_{\gamma = 1}^{N_3}
    e^{- i \bar{\phi}_3(\bar{w}_\gamma)}
    \right\rangle \nonumber \\
    &= \prod_\alpha (\eta - z_\alpha)^{\kappa_{11} v_1  + \kappa_{12} v_2 }
    \prod_\beta (\eta - u_\beta)^{\kappa_{21} v_1  + \kappa_{22} v_2}
    \prod_\beta (\bar{\eta} - \bar{u}_\beta)^{\bar{\kappa}_{22} v_2 + \bar{\kappa}_{23} v_3 } 
    \prod_{\gamma} (\bar{\eta} - \bar{w}_\gamma)^{\bar{\kappa}_{32} v_2 + \bar{\kappa}_{33} v_3} \Psi_1(z; u, w).
\end{align}
Equivalently, this can be expressed as
\begin{align}\label{eq:operator_insertion}
    \Psi_1(\eta; z) &= \left\langle e^{i v_1 \phi_1(\eta) + i v_2 [\phi_2(\eta) - \bar{\phi}_2(\bar{\eta})] -i v_3 \bar{\phi}_3(\bar{\eta})} \prod_{\alpha = 1}^{N_1}
    e^{i \phi_1(z_\alpha)}
    \prod_{\beta = 1}^{N_2}
    e^{i \phi_2(u_\beta) - i \bar{\phi}_2(\bar{u}_\beta)}
    \prod_{\gamma = 1}^{N_3}
    e^{- i \bar{\phi}_3(\bar{w}_\gamma)}
    \right\rangle \nonumber \\
    &= \prod_\alpha (\eta - z_\alpha)^{(\kappa \vec{v})_1}
    \prod_\beta (\eta - u_\beta)^{(\kappa \vec{v})_2}
    \prod_\gamma (\eta - w_\gamma)^{(\kappa \vec{v})_3} \nonumber \\
    &\quad \times \prod_\alpha (\bar{\eta} - \bar{z}_\alpha)^{(\bar{\kappa} \vec{v})_1}
    \prod_\beta (\bar{\eta} - \bar{u}_\beta)^{(\bar{\kappa} \vec{v})_2} 
    \prod_{\gamma} (\bar{\eta} - \bar{w}_\gamma)^{(\bar{\kappa} \vec{v})_3} \Psi_1(z; u, w).
\end{align}
In the plasma analogy, the first factor becomes
\begin{align}
    &\prod_{\alpha} (\eta - z_\alpha)^{(\kappa \vec{v})_1} = e^{(\kappa \vec{v})_1  \sum_\alpha \log (\eta - z_\alpha)} \nonumber \\
    &\to \exp \left( (\kappa \vec{v})_1 \int_{r'} \ln|\eta - r'| \rho_z(r') \right) 
    = \exp \left( (\kappa \vec{v})_1 \int_{r, r'} \delta^{(2)}(\eta - r) \ln|r - r'| \rho_z(r') \right) \nonumber \\
    &= \exp \left(\frac{1}{2} (\kappa \vec{v})_1 \int_{r, r'} \delta^{(2)}(\eta - r) \ln|r - r'| \rho_z(r') + \frac{1}{2} (\kappa \vec{v})_1 \int_{r, r'}  \rho_z(r) \ln|r - r'| \delta^{(2)}(\eta - r') \right).
\end{align}
Here, we retain only the real logarithmic part. The statistical phase factors can be removed by a singular-gauge transformation \cite{girvin1987off-diagonal}. However, for the null vectors of $M_K$ associated with the superfluid or superconducting states considered here, we have $\kappa \vec{v} = \bar{\kappa} \vec{v} = 0$. Consequently, the statistical phase due to the insertion vanishes identically. Therefore, omitting the phase factors does not affect the ODLRO argument in the cases of our interest. Similarly, the fourth factor in Eq.~\eqref{eq:operator_insertion} becomes
\begin{align}
    &\prod_\alpha (\bar{\eta} - \bar{z}_\alpha)^{(\bar{\kappa} \vec{v})_1} = e^{(\bar{\kappa} \vec{v})_1 \log(\bar{\eta} - \bar{z}_\alpha)} \nonumber \\
    &\to \exp \left((\bar{\kappa} \vec{v})_1 \int_{r'} \ln|\eta - r'| \rho_z(r') \right) \nonumber \\
    &= \exp \left(\frac{1}{2} (\bar{\kappa} \vec{v})_1 \int_{r, r'} \delta^{(2)}(\eta - r) \ln|r - r'| \rho_z(r') + \frac{1}{2} (\bar{\kappa} \vec{v})_1 \int_{r, r'}  \rho_z(r) \ln|r - r'| \delta^{(2)}(\eta - r') \right)
\end{align}
The remaining factors can be treated in the same way. Eq.~\eqref{eq:operator_insertion} can therefore be written as
\begin{align}
    &\exp \left(\frac{1}{2} \int_{r, r'} (\kappa \vec{v})^\mathsf{T} \delta^{(2)}(\eta - r) \vec{\rho}(r') \ln|r - r'| 
    + \frac{1}{2} \int_{r, r'} \vec{\rho}^\mathsf{T}(r) \kappa \vec{v} \delta^{(2)}(\eta - r') \ln|r - r'| \right) \nonumber \\
    &\times \exp \left(\frac{1}{2} \int_{r, r'} (\bar{\kappa} \vec{v})^\mathsf{T} \delta^{(2)}(\eta - r) \vec{\rho}(r') \ln|r - r'| + \frac{1}{2} \int_{r, r'} \vec{\rho}^\mathsf{T}(r) \bar{\kappa} \vec{v} \delta^{(2)}(\eta - r') \ln|r - r'| \right) \nonumber \\
    &=\exp \left(\frac{1}{2} \int_{r, r'} \frac{1}{2}(M_K \vec{v})^\mathsf{T} \delta^{(2)}(\eta - r) \vec{\rho}(r') \ln|r - r'| + \frac{1}{2} \int_{r, r'} \vec{\rho}(r) M_K \frac{1}{2}\vec{v} \delta^{(2)}(\eta - r') \ln|r - r'| \right) \nonumber,
\end{align}
where $\vec{\rho}(r) = (\rho_z (r), \rho_u (r), \rho_w (r))$. Here, we have used $\kappa + \bar{\kappa} = \frac{1}{2} M_K$ and suppressed $\Psi_1(z;u,w)$. Since $M_K^\mathsf{T} = M_K$, this can be rewritten as
\begin{align}
    &\Psi_1(\eta; z) \nonumber \\
    &= \exp \left(\frac{1}{2} \int_{r, r'} \left[\frac{1}{2}\vec{v} \delta^{(2)}(\eta - r) \right]^\mathsf{T} M_K \vec{\rho}(r') \ln|r - r'| + \frac{1}{2} \int_{r, r'} \vec{\rho}(r) M_K \left[\frac{1}{2}\vec{v} \delta^{(2)}(\eta - r')\right] \ln|r - r'| \right) \Psi_1(z; u, w).
\end{align}
We similarly obtain
\begin{align}
    &\Psi_1^*(\eta'; z) \nonumber \\
    &= \exp \left(\frac{1}{2} \int_{r, r'} \left[\frac{1}{2}\vec{v} \delta^{(2)}(\eta' - r) \right]^\mathsf{T} M_K \vec{\rho}(r') \ln|r - r'| + \frac{1}{2} \int_{r, r'} \vec{\rho}(r) M_K \left[\frac{1}{2}\vec{v} \delta^{(2)}(\eta' - r')\right] \ln|r - r'| \right) \Psi_1^*(z; u, w).
\end{align}
Therefore,
\begin{align*}
    &\Psi_1^*(\eta'; z) \Psi_1(\eta; z) \nonumber \\
    &= |\Psi_1(z; u, w)|^2 \nonumber \\
    &\quad \times  \exp \left(\frac{1}{2} \int_{r, r'} \left[\frac{1}{2}\vec{v} \delta^{(2)}(\eta' - r) \right]^\mathsf{T} M_K \vec{\rho}(r') \ln|r - r'| + \frac{1}{2} \int_{r, r'} \vec{\rho}(r) M_K \left[\frac{1}{2}\vec{v} \delta^{(2)}(\eta' - r')\right] \ln|r - r'| \right) \nonumber \\
    &\quad \times \exp \left(\frac{1}{2} \int_{r, r'} \left[\frac{1}{2}\vec{v} \delta^{(2)}(\eta - r) \right]^\mathsf{T} M_K \vec{\rho}(r') \ln|r - r'| + \frac{1}{2} \int_{r, r'} \vec{\rho}(r) M_K \left[\frac{1}{2}\vec{v} \delta^{(2)}(\eta - r')\right] \ln|r - r'| \right) 
\end{align*}
We define the effective action
\begin{equation}
    \mathcal{S}[\vec{\rho}] \equiv - \frac{1}{2} \int_{r,r'} \vec{\rho}(r)^\mathsf{T} M_K \vec{\rho}(r') \ln |r- r'|.
\end{equation}
Then $|\Psi_1(z; u, w)|^2 = e^{-\mathcal{S}[\vec{\rho}]}$, so that
\begin{align}
    &\Psi_1^* (\eta'; z) \Psi_1(\eta; z) \\
    &= e^{-\mathcal{S}[\vec{\rho} + \vec{j}_{\vec{v}}(\eta') + \vec{j}_{\vec{v}}(\eta)]} \nonumber \\
    &\quad \times 
    \exp \left(
    -\frac{1}{2} \int_{r, r'} 
    \left[\frac{1}{2}\vec{v} \delta^{(2)}(\eta' - r) \right]^\mathsf{T}
    M_K 
    \left[\frac{1}{2}\vec{v} \delta^{(2)}(\eta - r') \right]
    \ln|r - r'| 
    \right) \nonumber \\
    &\quad \times 
    \exp \left(
    -\frac{1}{2} \int_{r, r'} 
    \left[\frac{1}{2}\vec{v} \delta^{(2)}(\eta - r) \right]^\mathsf{T}
    M_K 
    \left[\frac{1}{2}\vec{v} \delta^{(2)}(\eta' - r') \right]
    \ln|r - r'| 
    \right) \nonumber \\
    &= e^{-\mathcal{S}[\vec{\rho} + \vec{j}_{\vec{v}}(\eta') + \vec{j}_{\vec{v}}(\eta)]} |\eta' - \eta|^{-\frac{1}{4} \vec{v}^\mathsf{T} M_K \vec{v}}.
\end{align}
Here, $\vec{j}_{\vec{v}}(\eta) = \frac{1}{2} \vec{v} \delta^{(2)}(\eta - r)$, and we have omitted the regulator-dependent self-interaction terms associated with $\eta-\eta$ and $\eta'-\eta'$. The first factor $e^{-\mathcal{S}[\vec{\rho} + \vec{j}_{\vec{v}}(\eta') + \vec{j}_{\vec{v}}(\eta)]}$ describes the fully interacting system of particles labeled by $\{\eta, \eta', z_1, \ldots, z_N\}$, including two external test charges at $\eta$ and $\eta'$. Therefore, the reduced density matrix is
\begin{equation}
    \rho(\eta, \eta') = \frac{\int_{z} \Psi_1^*(\eta'; z) \Psi_1(\eta; z)}{\int_z \Psi_1^*(z) \Psi_1(z)} 
    = 
    \frac{\int_{z} e^{-\mathcal{S}[\vec{\rho} + \vec{j}_{\vec{v}}(\eta') + \vec{j}_{\vec{v}}(\eta)]}}{\int_z e^{-\mathcal{S}[\vec{\rho}]}} |\eta' - \eta|^{-\frac{1}{4} \vec{v}^\mathsf{T} M_K \vec{v}}.
\end{equation}
This expression has the same form as the reduced density matrix considered in \cite{girvin1987off-diagonal}. Invoking the plasma-screening argument in \cite{girvin1987off-diagonal}, we obtain
\begin{equation}
    \frac{\int_z e^{-\mathcal{S}[\vec{\rho} + \vec{j}_{\vec{v}}(\eta') + \vec{j}_{\vec{v}}(\eta)]}}{\int_z e^{-\mathcal{S}[\vec{\rho}]}} = \frac{Z_{\text{full}}(\eta, \eta')}{Z_0} \sim C, \qquad |\eta - \eta'| \to \infty
\end{equation}
for some nonzero constant $C$. Therefore, at large separations, the reduced density matrix behaves as
\begin{equation}
    \rho(\eta, \eta') \sim C |\eta' - \eta|^{-\frac{1}{4} \vec{v}^\mathsf{T} M_K \vec{v}}, \qquad |\eta - \eta'| \to \infty.
\end{equation}
In particular, for the null vector $\vec{v} = \vec{v}_0$ satisfying $M_K \vec{v}_0 = 0$, the algebraic exponent vanishes. Consequently, the reduced density matrix approaches a nonzero constant,
\begin{equation}
    \rho(\eta, \eta') \to C, \qquad |\eta - \eta'| \to \infty,
\end{equation}
demonstrating nonvanishing ODLRO.

\end{document}